\documentclass[11pt,psfig]{article}
\pdfoutput=1
\usepackage{graphicx}
\usepackage{subfigure,amssymb,amsmath,slashed}
\usepackage{cite,color,xcolor,url,cancel}
\usepackage{bm}
\usepackage{float}
\usepackage{multirow}
\usepackage{multicol}
\usepackage{mathtools}
\usepackage{jheppub}
\usepackage{multirow, makecell}

\newcommand{\be}{\begin{equation}}
\newcommand{\ee}{\end{equation}}
\newcommand{\bea}{\begin{eqnarray}}
\newcommand{\eea}{\end{eqnarray}}

\newcommand{\bse}{\begin{subequations}}
\newcommand{\ese}{\end{subequations}}
\newcommand{\tp}{T^\prime}
\newcommand{\nn}{\nonumber}

\newcommand{\zp}{Z^\prime}
\newcommand{\rp}{\rho^\prime}
\newcommand{\xp}{x^\prime}

\newcommand{\x}{\chi}
\newcommand{\mt}{\mu \tau}
\newcommand{\lmt}{L_\mu - L_\tau}

\newcommand{\eps}{\epsilon}

\title{Secluded Dark Sector and Muon $(g-2)$ in the Light of Fast Expanding Universe}

\author[a]{Sougata Ganguly,}
\affiliation[a]{School of Physical Sciences, Indian Association 
	for the Cultivation of Science,\\  
	2A \& 2B Raja S.C. Mullick Road, Jadavpur, 
	Kolkata 700 032, India}

\author[a]{Sourov Roy,}
\author[a]{Ananya Tapadar}

\emailAdd{tpsg4@iacs.res.in}
\emailAdd{tpsr@iacs.res.in}
\emailAdd{intat@iacs.res.in}

\abstract{
The lack of information before Big Bang Neucleosynthesis (BBN) allow us to assume the presence of a 
new species $\phi$ whose energy density redshifts as $a^{-(4+n)}$ where $n>0$ and $a$ is the scale factor.
This non-standard cosmological setup facilitates a larger portal coupling $(\epsilon)$
between the dark and the visible sectors even when the two sectors are not in thermal equilibrium.
Here, we have considered $U(1)_{L_\mu-L_\tau}\otimes U(1)_X$ gauge extension of the Standard Model (SM) and studied
different phases of the cosmological evolution of a thermally decoupled dark sector such as leak-in, 
freeze-in, reannihilation, and late-time annihilation in the presence of fast expansion. Due to the
tree level kinetic mixing between $U(1)_X$ and $U(1)_{L_\mu-L_\tau}$ gauge bosons, the dark sector couples with
the $\mu$ and $\tau$ flavored leptons of the SM. We show that in our scenario it is possible to reconcile the dark matter relic density
and muon $(g-2)$ anomaly. In particular, we show that for $2\times 10^{-4} \lesssim \epsilon \lesssim 10^{-3}$, 
$5.5{\rm MeV} \lesssim m_{Z^\prime} \lesssim 200{\rm MeV}$, $n=4$, and $1{\rm TeV} \lesssim m_\x \lesssim 10{\rm TeV}$ 
relic density constraint of dark matter, constraint from muon $(g-2)$ anomaly, and other cosmological, astrophysical constraints are satisfied.
}
\vspace{1cm}
\graphicspath{{figure/}}

\begin{document}
\maketitle
\section{Introduction}
The Standard Model (SM) of particle physics is an extremely successful theory to explain observable phenomena
at the microscopic level. Despite huge successful predictions of SM, there are some unsolved issues which cannot be
explained by SM. The existence of dark matter (DM), revealed from various cosmological and astrophysical observations
(such as measurement of cosmic microwave background(CMB) anisotropy by WMAP \cite{WMAP:2012nax},
Planck 2018 \cite{Planck:2018vyg}, galaxy rotation curve \cite{Sofue:2000jx}, bullet cluster
observation \cite{Clowe:2003tk,Clowe:2006eq} etc.), is one such unsolved issue which SM cannot explain. 
According to the latest data from Planck 2018 observations \cite{Planck:2018vyg}, the abundance of DM at the present epoch 
is $\Omega_{\rm {DM}} h^2 = 0.1200 \pm 0.0012$ \cite{Planck:2018vyg}. Even after decades of first observational evidences, the origin and nature of 
DM are still elusive. None of the SM particles is suitable as a DM candidate and thus the physics beyond SM is indispensable.

Weakly interacting massive particle (WIMP) has been a well motivated DM 
candidate for a long time \cite{Srednicki:1988ce, Gondolo:1990dk, Jungman:1995df, 
Feng:2010gw}. Inspite of having many attractive theoretical features, WIMP paradigm 
is not suitable to explain the negative results of direct, indirect as well as laboratory searches \cite{Roszkowski:2017nbc, Lin:2019uvt}. As a result, the 
allowed parameter space of the WIMP scenario has been stringently constrained.

On the other hand, recent measurement of the anomalous magnetic moment of muon by Fermilab \cite{Muong-2:2021ojo} reveals
that the tension between the theoretical prediction \cite{Czarnecki:2002nt,Melnikov:2003xd,Aoyama:2012wk,Gnendiger:2013pva,Kurz:2014wya,Colangelo:2014qya,Masjuan:2017tvw,Colangelo:2017fiz,Davier:2017zfy,Keshavarzi:2018mgv,Hoferichter:2018kwz,Colangelo:2018mtw,Aoyama:2019ryr,Gerardin:2019vio,Hoferichter:2019mqg,Bijnens:2019ghy,Davier:2019can,Colangelo:2019uex,Keshavarzi:2019abf,Blum:2019ugy,Aoyama:2020ynm} and experimental observations of muon $(g-2)$
is at $4.2\sigma$. Combining the result of Fermilab with the old measurement by BNL E821 experiment \cite{Muong-2:2006rrc}, the deviation
of $a_\mu (\equiv (g-2)/2)$ from the SM prediction is given by \cite{Muong-2:2021ojo}
\bea
\Delta a_\mu &=& \left(251 \pm 59\right)\times 10^{-11} \nn\,\,.
\eea
Anomaly free $U(1)_{\lmt}$ gauge extension of the SM \cite{He:1990pn,He:1991qd} is a very well motivated model to explain
the mismatch between theoretical prediction and experimental observation of muon $(g-2)$ \cite{Baek:2001kca, Harigaya:2013twa, Altmannshofer:2016brv}.
$U(1)_{\lmt}$ gauge extension has also been extensively studied in the Supersymmetric framework
to explain the anomalous magnetic moment of muon \cite{Ma:2001md, Banerjee:2018eaf, Banerjee:2020zvi}. In the context of neutrino masses and mixings
$U(1)_{\lmt}$ model has been studied in \cite{Ma:2001md,Baek:2015mna,Chen:2017gvf}. 

In the light of the above discussion, we have considered $U(1)_{\lmt} \otimes  U(1)_X$ gauge extension of SM
to study the cosmological evolution of a thermally decoupled dark sector and muon $(g-2)$ anomaly. The dark sector
contains the gauge boson corresponding to the $U(1)_X $ gauge symmetry and the DM $\x$ which is a Dirac fermion, 
charged only under $U(1)_X$ gauge symmetry.  In order to connect dark and SM sector, we have considered tree level 
kinetic mixing between $U(1)_{\lmt}$ and $U(1)_X$ gauge bosons and the dark sector particles couple with the $\mu$ and $\tau$
flavored SM leptons through the portal coupling $\eps$. Let us note that, other possible tree level kinetic mixings between 
$U(1)_Y$ with $U(1)_{\lmt}$ and $U(1)_X$ gauge bosons are not considered here. However, these kinetic mixings can be generated
at one loop level. 

For a thermally decoupled dark sector, there exists an upper limit on the portal coupling $\eps$ and the temperature
evolution of both the sectors are very much different \cite{hambye:2019,Evans:2019vxr,Tapadar:2021kgw}. Since both the sectors
are thermally decoupled, the value of $\eps$ is very small. Due to the small value of the portal coupling, although the dark vector boson $\zp$,
having a coupling with muons, cannot explain muon $(g-2)$ anomaly. However, the situation
may improve if we modify the cosmological history before Big Bang Neucleosynthesis (BBN). The modification of the cosmology
in the pre-BBN era helps ameliorate the theoretical prediction of muon $(g-2)$ in the following manner.
If we assume the existence of a new species $\phi$ whose energy density redshifts as $a^{-(4+n)}$ where $n>0$ before BBN, then the Universe
expands faster in comparison to the standard radiation dominated Universe \cite{DEramo:2017gpl,DEramo:2017ecx}. The key
feature of the fast expanding scenario is the requirement of large coupling constant to satisfy the relic
density constraint. 
Therefore, to study muon $(g-2)$ anomaly and phases of a thermally decoupled dark sector in a unified framework, 
we have utilised the idea of fast expanding Universe to uplift the upper limit of the portal coupling $\eps$. In addition to that, 
we have also investigated the constraints from  Borexino \cite{Bellini:2011rx}, BABAR experiments \cite{BaBar:2016sci}, BBN, white dwarf cooling, self interaction of DM,and neutrino trident production at CCFR \cite{CCFR:1991lpl}.

The rest of the paper is structured as follows. In section \ref{Model} we have discussed our model briefly. 
The fast expanding scenario, thermalization and temperature evolution of the dark sector in the
fast expanding Universe has been discussed in section \ref{section-fast-expanding-univ}.
Analytical calculations for the DM relic density in fast expanding Universe has been performed in section \ref{properties-nonadiabetic-evl}.
We provide a detail analysis of DM number density evolution in section \ref{Numerical-result-for-relic-density}.
Consequence of $\zp$ on muon anomalous magnetic moment, relevant constraints on portal coupling
and dark vector boson mass are discussed in section \ref{constraints-on-lmt}. Finally, we conclude in section \ref{conclusion}.  
In appendix \ref{Appx-A}, we show the detailed analytical calculations of freeze-in scenario in case of fast expanding Universe.

\section{The model}
\label{Model}
We consider $U(1)_{\lmt} \otimes U(1)_X$ gauge extension of the SM to study the cosmological evolution of the
dark sector and muon $(g-2)$ anomaly simultaneously. We introduce a Dirac fermion $\x$ which is charged
only under $U(1)_X$ gauge symmetry. In our framework, $\x$ acts as a DM candidate and it only interacts with
the gauge boson of $U(1)_X$ gauge symmetry $\hat{\zp}$. All the SM fields are singlet under $U(1)_X$ gauge symmetry. 
We have considered the tree level kinetic mixing between $\hat{\zp}$ and $U(1)_{\lmt}$ gauge boson $\hat{Z}_{\mt}$ and because of this, 
the dark sector couples with the $\mu$ and $\tau$ flavored SM leptons. One of the interesting features of this model 
is that the dark gauge boson i.e. the gauge boson corresponding to the $U(1)_X$ gauge symmetry can contribute to 
the anomalous magnetic moment of muon if it is sufficiently light. In our model we have assumed that the masses of the gauge
bosons are generated via St{\" u}ckelberg mechanism \cite{Stueckelberg:1938hvi, Ruegg:2003ps}. As mentioned in the introduction, 
tree level kinetic mixing between $U(1)_Y$ gauge boson with $U(1)_{\lmt}$ and 
$U(1)_X$ gauge bosons are neglected here. However, we would like to mention that these mixings can be generated at one loop level
and we have considered the effect of loop induced kinetic mixing in our discussion on the constraints of dark gauge boson.

The Lagrangian of our model is given by \cite{Tapadar:2021kgw}
\bea
\label{Lagrangian}
\mathcal{L}
&=&
\mathcal{L}_{\rm SM}  + \bar{\x}\left(i \slashed{\partial} - m_\x\right)\x
- \dfrac{1}{4} \hat {X}^{\rho \sigma} \hat{X}_{\rho \sigma}
- \dfrac{1}{4} \hat {F}_{\mt}^{\rho \sigma} \hat{F}_{\mt_{\rho \sigma}}
+ \dfrac{\sin \delta}{2} \hat{F}_{\mu\tau}^{\rho \sigma} \hat{X}_{\rho \sigma}
\nn\\
&&
- g_X \bar{\x} \gamma^\rho \x {\hat{\zp}}_\rho
-g_{\mu \tau}
\left(
\bar{\mu}\gamma_\rho \mu + \bar{\nu}_\mu \gamma_\rho P_L \nu_\mu
-\bar{\tau}\gamma_\rho \tau - \bar{\nu}_\tau \gamma_\rho P_L \nu_\tau
\right)\hat{Z}_{\mt}^\rho\nn\\
&&
+\dfrac{1}{2} \hat{m}_{\mt}^2 \hat{Z}_{\mt}^\rho 
\hat{Z}_{\mt_\rho}
+\dfrac{1}{2} \hat{m}^{\prime 2} \hat{\zp}^\rho \hat{\zp}_\rho \,\,.
\label{eq:Lag1}
\eea
Here, $\hat{X}^{\rho\sigma} = 
\partial^\rho\hat{\zp}^\sigma - \partial^\sigma \hat\zp^\rho$ and $\hat{F}_{\mt}^{\rho \sigma} = \partial^\rho \hat{Z}_{\mt}^\sigma - 
\partial^\sigma \hat{Z}_{\mt}^\rho$ are the field strength tensors of $U(1)_X$ and $U(1)_{\lmt}$ gauge symmetry respectively.
$m_\x$ is the mass of the DM $\x$, $\hat{m}^\prime$ and $\hat{m}_{\mt}$ are the gauge bosons' mass parameters of the theory. The gauge couplings
corresponding to $U(1)_X$ and $U(1)_{\lmt}$ are denoted by $g_X$ and $g_{\mt}$ respectively. The last term in the first line
of Eq.\,\ref{eq:Lag1} denotes the tree level kinetic mixing between $\hat{Z}^\prime$ and $\hat{Z}_{\mt}$. 

To put the Lagrangian in canonical form,  we have performed a non-orthogonal 
transformation from ``hatted" to ``tilde" basis and the transformation is
\bea
\label{non-ortho-trans_1}
\begin{pmatrix}
	\hat{Z}^\prime_\rho \\
	\hat{Z}_{\mt_\rho}
\end{pmatrix}
=
\begin{pmatrix}
	\sec \delta & 0\\
	\tan \delta & 1
\end{pmatrix}
\begin{pmatrix}
	\tilde{\zp}_\rho\\
	\tilde{Z}_{\mt_\rho}
\end{pmatrix}\,\,.
\eea
Removal of kinetic mixing term generates mass mixing between $\tilde{\zp}$ and 
$\tilde{Z}_{\mu\tau}$ bosons. This can be removed by transforming ``tilde" 
basis into mass basis by the following orthogonal transformation
\bea
\label{ortho-trans}
\begin{pmatrix}
\tilde{Z}_{\mt_\rho}\\
\tilde{Z}^\prime_\rho
\end{pmatrix}
&=&
\begin{pmatrix}
	\cos \beta & -\sin \beta\\
	\sin \beta & \cos \beta
\end{pmatrix}
\begin{pmatrix}
	Z_{\mt_\rho}\\
	\zp_\rho
\end{pmatrix}
\,\,\,,
\eea
where $\beta$ is the $\zp -Z_{\mt} $ mixing angle and in the limit $\delta \ll 1$, it is given by
\bea
\tan \beta & \simeq & \dfrac{\tan \delta}{1- \hat{r}^2}\,\,,\
\label{mixing-angle}
\eea
where $\hat{r} =  m^{\prime 2}/ m_{\mt}^2$, $m^\prime$ and $m_{\mt}$ are the masses of $\zp$ and $Z_{\mt}$ respectively. In the limit $\delta,\,\hat{r} \ll 1$,
$\hat{Z}$ and $\hat{Z}_{\mt}$ can be expressed in the following manner
\bea
\hat{Z}_{\mt_\rho} &\simeq& Z_{\mt_\rho} -\hat{\eps} \zp_\rho\,\,,\nn\\
\hat{Z}^\prime_\rho &\simeq& \zp_\rho + \tan \delta Z_{\mt_\rho} \,\,,
\label{gauge-boson}
\eea
where $\hat{\eps} = \hat{r}^2 \rm{tan} \, \delta $. Using Eq.\,\ref{gauge-boson}  
we can rewrite the interaction term of dark vector boson $Z^\prime$ as follows
\bea
\mathcal{L}
\label{Lagrangian_2}
\supset
-g_X \bar{\x} \gamma^\rho \x \zp_\rho
+\eps\left(
\bar{\mu}\gamma^\rho \mu + \bar{\nu}_\mu \gamma^\rho P_L \nu_\mu
-\bar{\tau}\gamma^\rho \tau - \bar{\nu}_\tau \gamma^\rho P_L \nu_\tau
\right)\zp_\rho\,\,\,.
\label{eq:Lag2}
\eea
Here, $\eps = g_{\mu\tau}\hat{\eps}$ is the portal coupling between dark and the visible sector. Note that the condition $\hat{r}\ll 1$
tells us that $Z_{\mt}$ will have negligible contribution to muon $(g-2)$ anomaly and we have only considered the contribution of $\zp$ to the muon
$(g-2)$ anomaly in the rest of our discussion.

\section{Fast expanding Universe}
\label{section-fast-expanding-univ}
\subsection{Cosmology of fast expanding Universe}
\label{fast-expanding-universe}
Prediction of light element abundance at the time of BBN tells us that our Universe was radiation dominated at the time of BBN.
However, it is possible that the Universe was not radiation dominated before BBN. This implies that a new species $\phi$ might have existed at the 
pre-BBN era whose energy density redshifts faster than radiation \cite{DEramo:2017gpl,DEramo:2017ecx}. In that case, total energy density of 
the Universe can be expressed as
\bea
\rho(T) &=& \rho_\phi(T) + \rho_{r}(T)+ \rho_{r}^\prime(\tp)\,\,,
\label{total_energy}
\eea
where $\rho_{r}(T)  = \dfrac{\pi^2}{30}g_\rho(T) T^4$  is the radiation energy density of the SM bath
at SM temperature $T$, $\rho_{r}^\prime(\tp) = \dfrac{\pi^2}{30}g_{\rho^\prime}(\tp) {\tp}^4 $ is the radiation energy density 
of dark sector (DS) and $\tp$ is the temperature of the DS. $\rho_\phi (T)$ is the energy density of the new species $\phi$ and it
redshifts as $a^{-(4+n)}$ where $a$ is the scale factor and $n>0$. Here, $g_\rho(T)$ and $g_{\rho^\prime}(\tp)$ 
are effective number of relativistic degrees of freedom of SM sector and DS respectively, contributing to the energy density. 
Here, we have assumed $\tp \ll T$ and we may neglect DS sector energy density compared to SM bath. Therefore the total energy density can be expressed as
\bea
\rho(T) &=& \rho_\phi(T) + \rho_{r}(T).
\label{eq:total_energy_density}
\eea

Motivated by the constraints from BBN, one can define a temperature $T_r$ at which\footnote{The abundances of the light elements
at the time of BBN remain unaltered if we choose $T_{r} \gtrsim (15.4)^{1/n} \rm MeV$ where $n>0$ \cite{DEramo:2017gpl}.}
$\rho_{\phi} (T_r)= \rho_r (T_r)$. Using this
definition of $T_r$ and entropy conservation of the Universe, we can write $\rho_{\phi} (T)$ as
\bea
\rho_\phi(T) &=& \rho_r(T)\left(\dfrac{g_\rho(T_{r})}{g_\rho(T)}\right)
\left(\dfrac{g_{*s}(T)}{g_{*s}(T_{r})}\right)^{\frac{4+n}{3}}\left(\dfrac{T}{T_{r}}\right)^{n}\,\,.\,
\label{eq:E_phi_T} 
\eea
Therefore, total energy density can be written as
\bea
\rho(T) &=& \rho_{r}(T)\left[1+ \left(\dfrac{g_\rho(T_{r})}{g_\rho(T)}\right)
\left(\dfrac{g_{*s}(T)}{g_{*s}(T_{r})}\right)^\frac{4+n}{3}
\left(\dfrac{T}{T_{r}}\right)^{n}
\right] \,\,.\,
\label{eq:E_total_T}
\eea
It is clear from Eq.\,\ref{eq:E_total_T} that at  early times, i.e. $T \gg T_{r}$, 
energy  density of $\phi$ dominates the energy budget of the Universe and the Universe expands 
faster than radiation dominated era. Similarly, for $T \ll T_{r}$, one can see $\rho(T) \simeq \rho_{r}(T)$.
We would like to mention that throughout our analysis we have chosen $T_r=20\rm MeV$ and the choice
of $T_r$ is consistent with the BBN constraint as discussed earlier.

Now, in presence of a new species $\phi$, Hubble parameter is given by
\bea
H(T) &=& \sqrt{\dfrac{8\pi}{3 M_{\rm Pl}^2}\rho(T)}\,\,,
\label{eq:hubble}
\eea
where $ M_{\rm{Pl}} = 1.22 \times 10^{19} \rm GeV$ is the Planck mass.
At high temperature limit  i.e. $T \gg T_{r}$, $H(T)$ can be approximated as
\bea
H(T) &\simeq& \dfrac{\pi}{3 M_{\rm Pl}}\sqrt{\dfrac{4 \pi}{5}} 
\sqrt{g_\rho(T)} T^2 \left(\dfrac{T}{T_{r}}\right)^{n/2}\,\,.\
\label{eq:hubble_high}
\eea

\subsection{Thermalisation of visible sector and dark sector}
\label{wimp-next-door}
\begin{figure}
\centering
\includegraphics[height = 10cm, width = 12cm]{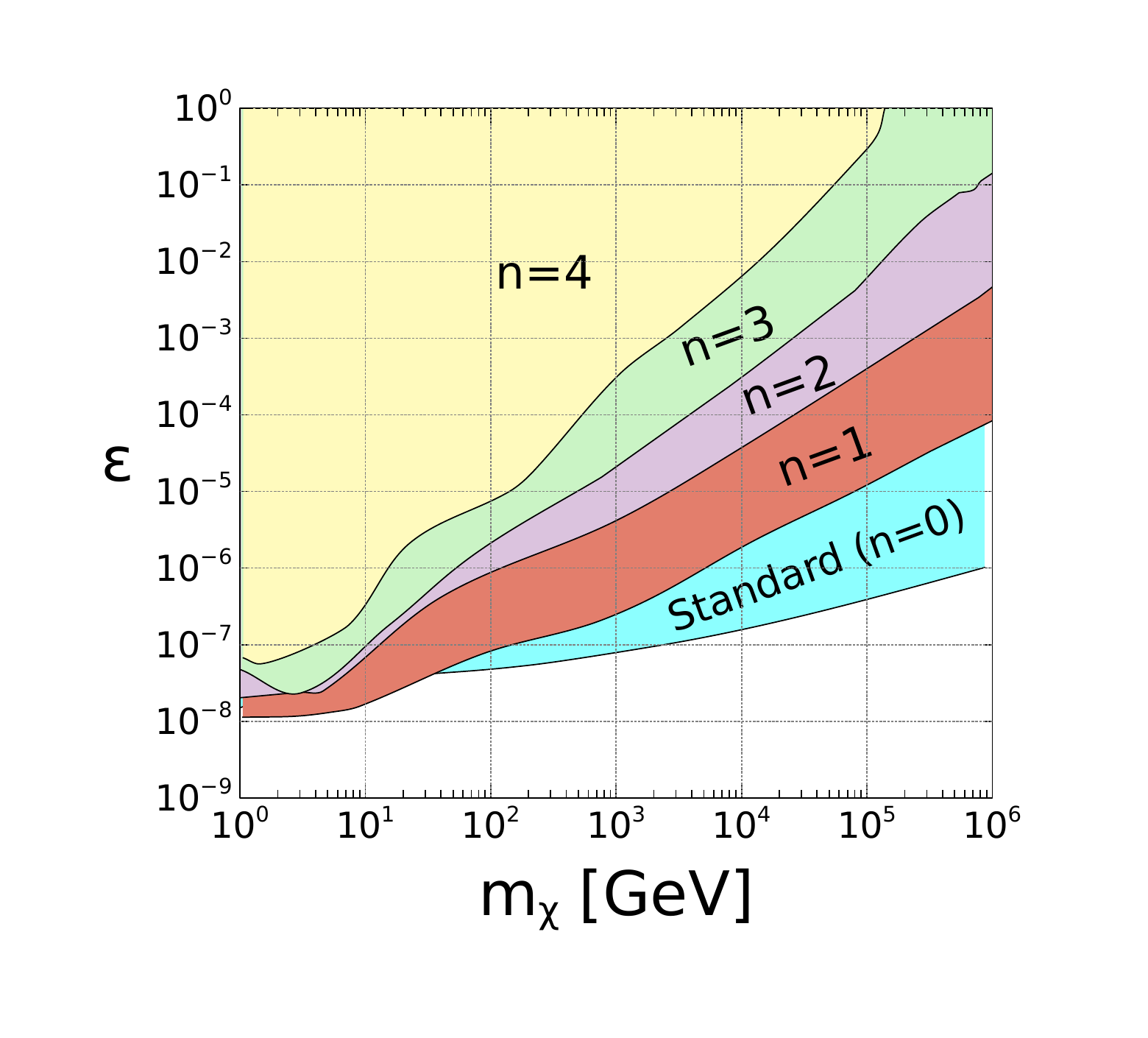}
\caption{
Allowed regions in $m_\x-\eps$ plane from the thermalisation criterion for $n=1$ (red),  $n=2$ (violet),  $n=3$ (green),  $n=4$ (yellow)
are shown. The constraint considering radiation dominated Universe is shown with cyan color.}
\label{wimp}
\end{figure}
Thermalisation of DS and SM bath not only 
depends on the interaction strength of the portal coupling, but also on the expansion rate 
of the Universe. In standard radiation dominated Universe for large portal coupling, there would be sufficient amount of energy exchange 
between the two sectors and they will be in thermal equilibrium with each other. 
In this framework DM freezes-out as the universe expands and DM relic abundance is determined by its annihilation into dark vector boson, 
$\x\bar{\x}\to \zp\zp$ if $m_\x > m_{\zp} $. This is known as secluded sector freeze out \cite{ Pospelov:2007mp, Feng:2008mu,
Cheung:2010gj,Chu:2011be,Evans:2017kti}.

In our study, the central idea of DM number density evolution is focused on the fact 
that two sectors are not in thermal equilibrium with each other. Therefore, first we identify 
the parameter space in $m_\x-\eps$ plane in which the dark sector is not in thermal equilibrium with the visible sector. 
To identify  this region,  initially we assume that the two sectors are in 
thermal equilibrium. We calculate reaction rate $\left(\Gamma\right)$ of all processes producing $\zp$ from SM bath (the processes
are mentioned in section \ref{dark-sector-temperature}). If $\Gamma(T)_{{\rm SM} \rightarrow \zp}\gtrsim H(T)$ at the time of DM freeze-out,  then two 
sectors are in thermal contact with each other, at the time of DM freeze-out. This region is shown in Fig.\ref{wimp} for different values of 
expansion parameter $n$ which controls the expansion rate of the Universe. 

In this scenario, at early Universe in the absence of any mass thresholds, reaction rates go as
$\Gamma_{{\rm SM}\rightarrow\zp}\sim \eps^2 T$. Now, in case of radiation dominated Universe, the ratio
$\Gamma_{{\rm SM} \rightarrow\zp}/H(T)$, at the time of DM freeze-out, goes as $\sim \eps^2/m_\x$. 
Thus we need large value of $\eps$ as we increase the value of $m_\x$ for the dark sector to be thermalised with the SM bath 
and we can see this from Fig.\,\ref{wimp}. However in case of fast expanding Universe, $H(T)$ is much larger in comparison to the 
radiation dominated Universe as one can see from Eq.\,\ref{eq:hubble_high}. As a result, the required value of $\eps$ to establish thermal equilibrium
between dark and visible sectors will be larger in comparison to the radiation dominated Universe. Thus for a fixed value of $m_\x$, the upper limit
of $\eps$ for thermally decoupled dark sector increases as we increase the value of $n$. In Fig.\,\ref{wimp}, we show the
parameter space in $m_\x-\eps$ plane for thermally decoupled dark sector and one can clearly see from the figure that
the required value of $\eps$ for the thermalisation of dark sector increases with the increase in the value of $n$.

Modification of the cosmological history of the early Universe opens up the possibility of explaining muon $(g-2)$ anomaly while
the dark sector is still thermally decoupled. In particular, for $\eps \sim 10^{-3}$, the dark sector is still thermally
decoupled if $n\geq3$ and $m_\x \gtrsim 30 \rm TeV$. However, in this region of parameter space,  the dark gauge boson $\zp$ can elevate 
the theoretical prediction of muon $(g-2)$ if $m_{\zp} \sim \mathcal{O}(100\rm MeV)$. In the subsequent sections, we will discuss
the DM production mechanisms for the thermally decoupled dark sector in fast expanding Universe.
\subsection{Temperature evolution of dark sector} 
\label{dark-sector-temperature}

To study the temperature evolution of the DS,  our next step is to calculate DS temperature $\tp$. In our
scenario, dark sector is populated from $\mu$ and $\tau$ flavored SM leptons involving annihilations processes in the early Universe and the DS
temperature depends on interaction strength of the portal 
coupling $(\eps)$ between DS and VS, $T_r$, and $n$. To calculate $\tp$, we have considered all
possible $\zp$ production
processes from SM bath such as 
$l\left(\bar{l}\right) + W^{+}\left(W^{-}\right) \rightarrow \nu_l \left(\bar{\nu_l}\right) + \zp  $,\,
$\nu_l \left(\bar{\nu_l}\right) + W^{-}\left(W^{+}\right) \rightarrow l \left(\bar{l}\right) +  \zp  $,\,
$l\left(\bar{l}\right) + \nu_l \left(\bar{\nu_l}\right) \rightarrow W^{-}\left(W^{+}\right) + \zp $,\,
$l+\bar{l} \rightarrow \gamma\left(Z,h\right) + \zp$,\,
$\nu_l+\bar{\nu_l} \rightarrow Z\left(h\right) + \zp$,\,
$l + \gamma\left(Z,h\right) \rightarrow l + \zp  $,\,
$\bar{l} + \gamma\left(Z,h\right) \rightarrow   \bar{l} + \zp  $,\,
$\nu_l + Z\left(h\right) \rightarrow  \nu_l + \zp  $,\,
$\bar{\nu_l} + Z\left(h\right) \rightarrow  \bar{\nu_l} + \zp $,\,
$l + \bar{l} \rightarrow \zp$, $\nu_l+\bar{\nu_l} \rightarrow \zp$
where, $l = \lbrace\mu,\tau \rbrace$. 
Now the DS temperature evolution can be obtained from the Boltzmann equation for the DS energy density $\rho^\prime$, and it is given by,
\bea
\label{BE_energy}
\dfrac{d \rho^\prime}{dt} + 4 H \rho^\prime  &\simeq & \mathcal{C}_{{\rm SM}\to \zp} \left(T\right)\,\,,\
\eea
where $\mathcal{C}_{{\rm SM}\to \zp}$  is the collision term for above-mentioned processes. In our calculation of the collision term, 
we have only considered the processes which are proportional to $\eps$ at the amplitude level.
Since $\tp \ll T$, energy exchange from DS to VS is very small, and therefore we have not considered the collision term 
due to dark sector annihilation into SM. Now, we have defined
$\xi$ as $\xi \equiv \dfrac{\tp}{T}$ and using Eq.\,\ref{BE_energy}, the solution of $\xi (T)$ as a function of $T$ is as follows
\bea
\xi (T)
&=&
\left[\int_{T}^{T_0}
\dfrac{30 \,\mathcal{C}_{{\rm SM}\to \zp} (\tilde{T})}
{g_{\rp} \pi^2 H(\tilde{T}) \tilde{T}^5}
d \tilde{T}\right]^{1/4}\,\,.\,
\label{energy_BE}
\eea
Here, $T_0$ is the initial temperature of the Universe
and we take $\tp \simeq 0$ at $T= T_0$,  $g_{\rp}$ is the relativistic degrees of
freedom contributing to the radiation bath of the
dark sector, $ H(T)$ is the expansion rate of the Universe as given in Eq.\,\ref{eq:hubble}. 
\begin{figure}
\centering
\includegraphics[height = 10cm, width = 12cm]{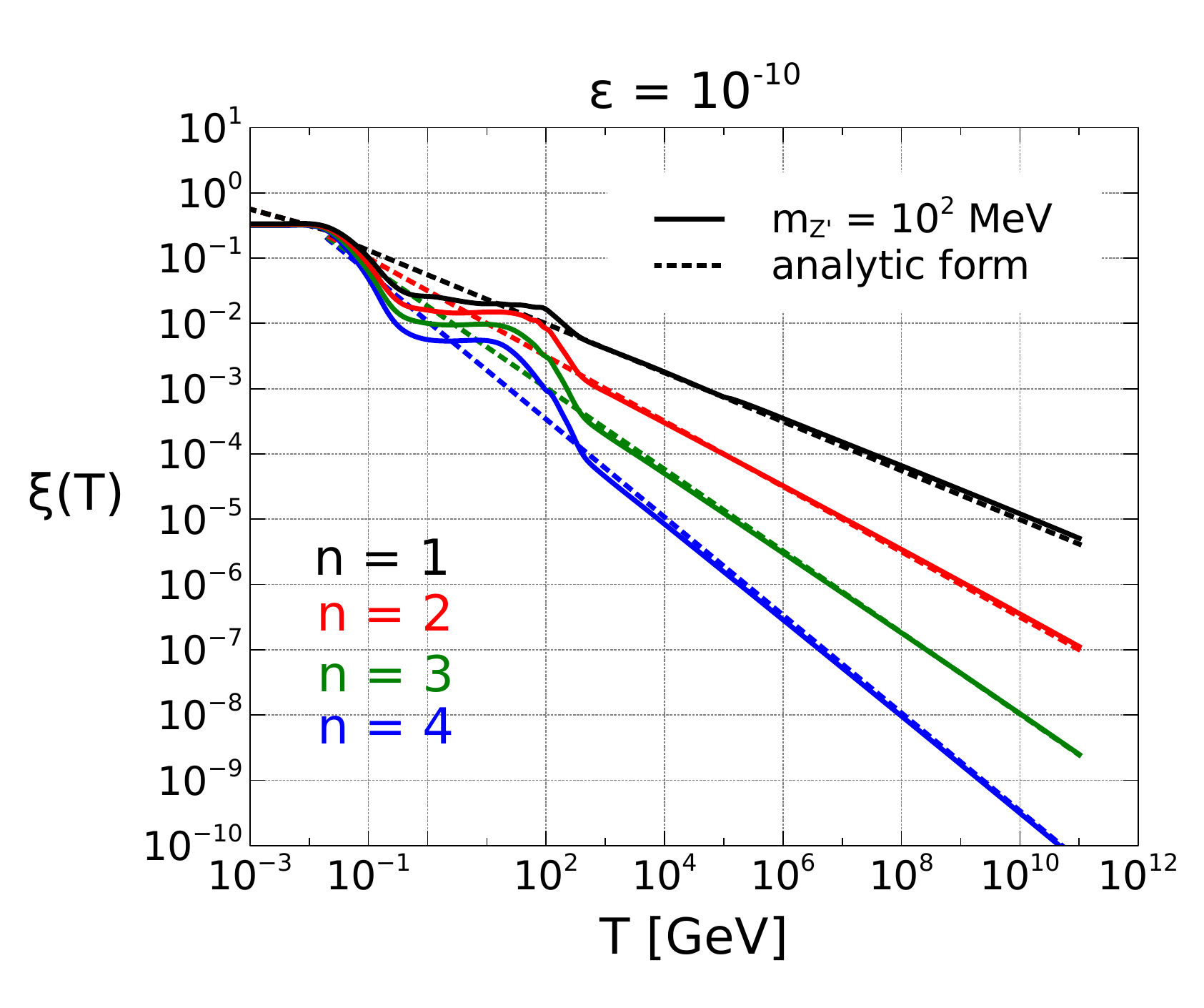}
\caption{
Evolution of $\xi(T)$ (solid lines) as a function of $T$ for $n=1$(black) for $n=2$ (red), $n=3$ (green), $n=4$ (blue).
The dashed lines with same color codes denote the behavior of semi-analytic expression of $\xi(T)$ as a function of $T$
(given in Eq.\,\ref{Tprime}). Here, we have considered $m_{\zp} = 100\rm MeV$ and $\eps=10^{-10}$.
}
\label{xi_vs_T}
\end{figure}

In Fig.\,\ref{xi_vs_T}, we show the variation of $\xi(T)$ as a function of $T$ for $m_{\zp}=100\rm MeV$  and $\eps=10^{-10}$
for four different values of $n$. One can clearly see that at $T\sim m_{\zp}$ energy injection stops and after that $\xi(T)$
remains constant. Another important feature is that the value of $\xi(T)$ at a fixed value of $T$ decreases as we increase $n$.
This is because with the increase in $n$, the energy injection processes decouple earlier and as a result amount of energy
injection will be less. 

\begin{figure}
\centering
\includegraphics[height = 10cm, width = 12cm]{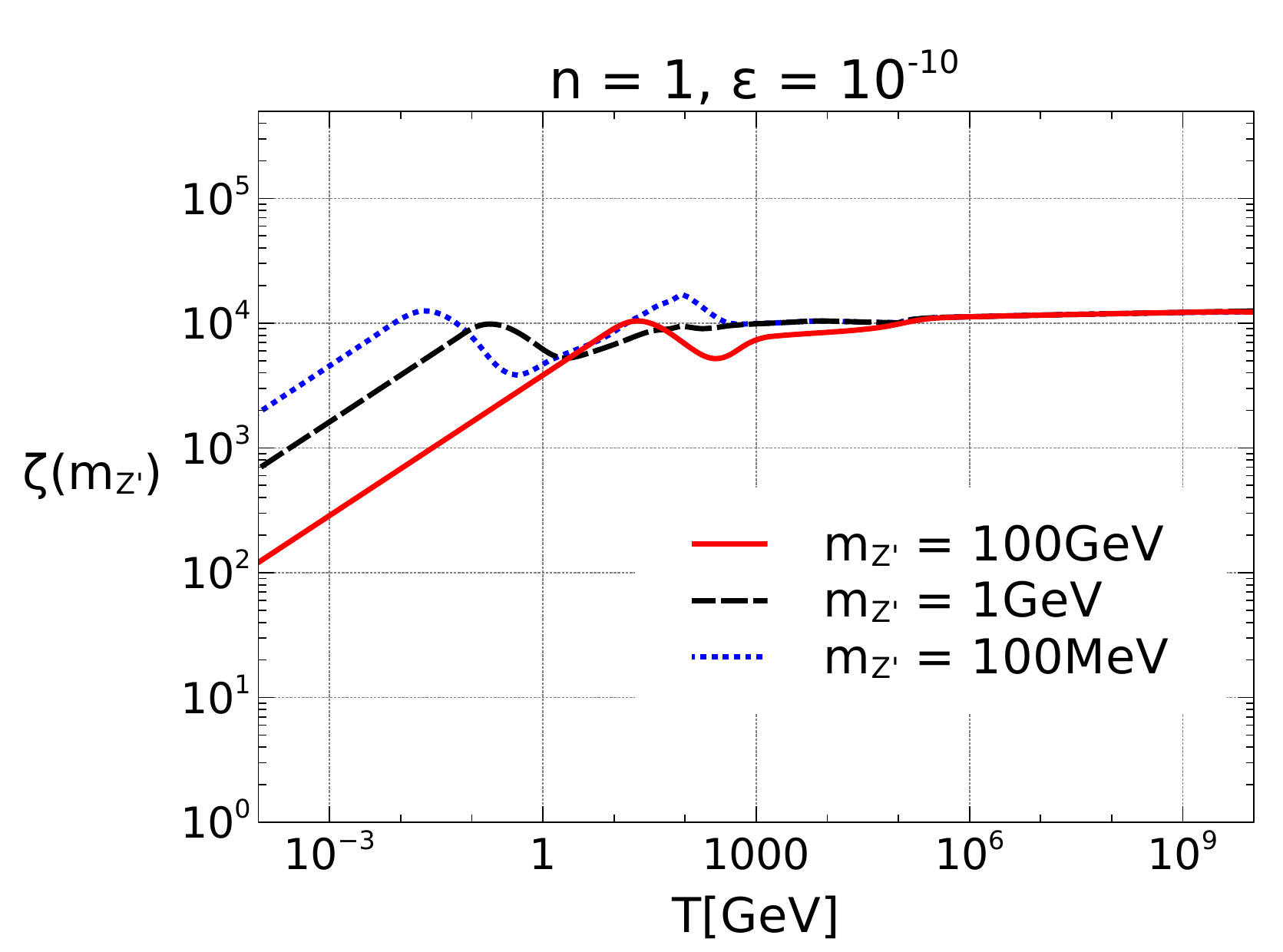}
\caption{
Variation of $\zeta (m_{\zp})$ as a function $T$ for $m_{\zp} =100\rm MeV$ (blue dotted line),
$1\rm GeV$ (black dashed line), and $100\rm GeV$ (red solid line). Here we have considered $\eps = 10^{-10}$
and $n=1$.}
\label{zeta_vs_T}
\end{figure}

In our scenario, the collision terms are proportional to the portal coupling $\eps^2$ and the portal interaction is renormalisable. 
Since the portal interaction is renormalisable, using Eq.\,\ref{energy_BE} one can write a semi-analytic 
expression for $\xi$ as follows \cite{Evans:2019vxr, Tapadar:2021kgw}
\bea
\label{Tprime}
\xi &\simeq& \dfrac{\zeta(m_{\zp}) \sqrt{\eps}}{(1+\frac{n}{2})^{\frac{1}{4}}}\,
T^{-\frac{1}{4}}\left(\dfrac{T_r}{T}\right)^{\frac{n}{8}}\,\, {\rm for} \,\, T > T_r \nn \\
&\simeq& \zeta(m_{\zp})\sqrt{\eps} T^{-\frac{1}{4}} \,\,\, {\rm for} \,\,  T \leq T_r,
\eea
where $\zeta (m_{\zp})$ will be extracted from our numerical analysis. Since the effect of non-standard cosmology
will enter into the evolution of $\xi$ only through Eq.\,\ref{eq:hubble}, therefore $\zeta(m_{\zp})$ does not depend on $n$.
In Fig.\,\ref{zeta_vs_T}, we show the variation of $\zeta (m_{\zp})$ for different values of $m_{\zp}$ and one can see
that for $T\gtrsim 10 m_{\zp}$, $\zeta(m_{\zp}) \simeq 10^4$. Thus during the phase of energy injection, one can
use Eq.\,\ref{Tprime} for the evolution of $\tp$. The dashed lines of Fig.\,\ref{xi_vs_T} denote
the behaviour of Eq.\,\ref{Tprime} as a function of $T$ for different values of $n$. 
\section{Properties of non-adiabatic evolution in fast expanding Universe}
\label{properties-nonadiabetic-evl}

Dark matter freeze-out through non-adiabatic evolution in fast expanding Universe 
is quite different from either of non-adiabatic evolution scenario as discussed 
earlier in \cite{Evans:2017kti,Tapadar:2021kgw} or dark matter freeze-out in 
non-standard cosmological scenario \cite{DEramo:2017gpl}. We shall start with some 
analytical calculation of DM freeze-out in this scenario and also put some bounds 
on portal coupling.

Dark matter freeze-out within DS through non-adiabatic evolution is referred to as  ``leak-in" 
dark matter \cite{Evans:2019vxr}. Difference between non-adiabatic evolution of  DM number density 
with dark sector freeze-out is that in the former scenario, $\xi$ is a dynamical quantity which
depends on the portal coupling as well as the SM temperature whereas $\xi$ is constant in the latter
scenario. Now, the relevant Boltzmann equation for dark sector freeze-out is given by
\bea
\dfrac{dY_{\x}}{dx} &=& \dfrac{\langle \sigma v \rangle^{\tp} s(x)}{x H(x)} 
\left({Y_{\x}^{\rm eq}}(\xp)^2 -Y_{\x}^2\right)\,\,,\
\label{BE-appx}
\eea
where, $x = m_\x/T$, $Y_{\x}^{\rm eq}(\xp) = n_{\x}^{\rm eq}(\xp)/s(x)$, where $n_{\x}^{\rm eq}(\xp)$ is the equilibrium DM number density, $s(x)$ is the
entropy density of the Universe, and $\xp = m_\x/\tp$. $\langle \sigma v \rangle^{\tp}$ is the thermally averaged cross-section of DM annihilation, 
calculated at $\tp$ and $H(x)$ is the Hubble parameter which is given in Eq.\,\ref{eq:hubble}.

Initially, DM comoving number density follows $Y_{\x}^{\rm eq}(\xp)$ and at later times it decouples 
from  dark sector thermal bath and freezes-out at $\xp_f = \dfrac{m_\x}{\tp_f}$ where, $\tp_f$ is the
dark sector temperature at the time of DM freeze-out. In case of fast expanding Universe, due to the higher expansion rate,
number changing processes for DM decouple much earlier in comparison to the radiation dominated Universe.
Therefore, in this scenario $x_f^\prime$ appeared to be smaller than the radiation dominated Universe. 
Exact value of $x_f^\prime$ can be obtained by solving the following equation \cite{DEramo:2017gpl}
\bea
\label{xp_f_exact}
\dfrac{e^{x^\prime_f}{\sqrt{x_f^\prime}}}{\xi \langle \sigma v \rangle}\left(\dfrac{x_r}{\xi x^\prime_f}\right)^{\frac{n}{2}} &=&
\dfrac{c\left(c+2\right)}{\left(c+1\right)}\dfrac{3\sqrt{5}}{4\sqrt{2}\pi^3}\dfrac{g_\x}{\sqrt{g_{*s}}}m_\x  M_{\rm{Pl}}\,\,,\
\eea 
where $g_\x$ is the degrees of freedom of DM, which is $2$ in our scenario, $x_r = m_\x/T_r$, 
$g_{*s}$ is effective degrees of freedom contributing to entropy density, and 
$c$ is a constant which is numerically found to be equal to $c(c+2)=\left(\frac{n}{2}+1\right)$ \cite{Scherrer:1985zt}. 
For $s$-wave DM annihilation cross-section Eq.\,\ref{xp_f_exact} can be approximated as,
\bea
\label{xp_f}
x^\prime_f \simeq \ln A -\dfrac{n}{2} \ln x_r\,\,,\
\eea 
where $A = \dfrac{c\left(c+2\right)}{\left(c+1\right)}\dfrac{3\sqrt{5}}{4\sqrt{2}\pi^3}\dfrac{g_\x}{\sqrt{g_{*s}}}\langle \sigma v \rangle m_\x M_{\rm Pl}$.

To get an analytical expression of final relic abundance $Y_\infty$, we have considered $s$-wave dominated annihilation cross section. 
As DM freeze-out temperature from DS thermal bath is $\tp_f$ then SM bath temperature at the time of DM freeze-out is $\tp_f/\xi$. 
At late time after freeze-out $Y_{\x}^{\rm eq}(\tp) \simeq 0$ compared to $Y_\infty$ which is the final value of DM comoving number density. 
Therefore, final comoving number density of DM is given by,
\bea
\label{relic-appx-eq}
\dfrac{1}{Y_\infty} &=& \dfrac{1}{Y_f\left(\xi x^\prime_f \right)} + \langle \sigma v \rangle_0 
\int_{\xi x^\prime_f}^{x_r} \dfrac{s(x)}{x H(x)} dx +\int_{x_r}^{\infty} \dfrac{s(x)}{x H_{\rm{rad}}(x)} dx\,\,,\
\eea  
where $ \langle \sigma v \rangle_0$ is the s-wave thermally averaged DM  annihilation cross section,
$H(x)$ is the Hubble parameter in fast expanding Universe as approximated in Eq.\,\ref{eq:hubble_high}. 
For $T<T_r$, Universe is radiation dominated and the Hubble parameter in this 
regime is defined by $H_{\rm{rad}}(x) = \dfrac{1}{M_{\rm Pl}}\sqrt{\dfrac{8\pi}{3}\rho_r(T)}$.

Therefore, by integrating  Eq.\,\ref{relic-appx-eq}, approximated expression for the comoving number density is given by,
{\small \bea
	\label{relic-appx}
	Y_\infty &=& \sqrt{\dfrac{45}{\pi}} \dfrac{\sqrt{g_\rho(x)}}{g_{*s}(x)} \dfrac{2 x_r}{\langle \sigma v \rangle m_\x M_{\rm Pl}} 
	\left[\dfrac{2}{\xi x^\prime_f}\left(\dfrac{\xi x^\prime_f}{x_r}\right)^{\frac{n}{2}-1}+
	\dfrac{1}{n-2}\left(1-\left(\dfrac{\xi x^\prime_f}{x_r}\right)^{\frac{n}{2}-1}\right) + 1\right]^{-1} \,,\ n\neq 2\nn\\
\\
	&=& \sqrt{\dfrac{45}{\pi}} \dfrac{\sqrt{g_\rho(x)}}{g_{*s}(x)} \dfrac{x_r}{\langle \sigma v \rangle m_\x {M_{\rm Pl}}} 
	\left[\dfrac{2}{\xi x^\prime_f}+ {\rm ln} \left(\frac{x_r}{\xi x^\prime_f} \right)+1\right]^{-1}\,,\ n=2 \,\,.\
	\eea}
Eq.\,\ref{relic-appx} reduces to standard WIMP
scenario as soon as we put $\xi = 1$ and $n=0$. Thus, it is clear that apart 
from the interactions within DS, final relic abundance for 
internally thermalized DS
in non-standard cosmology also depends on $n$, portal interaction 
between DS and VS and $T_r$.

DM production through this mechanism can take place within a certain range 
of portal coupling. Leak-in mechanism is only possible when dark and 
visible sectors are not in thermal equilibrium, but there is small exchange 
of energy between the two sectors. If portal coupling is sufficiently large 
then both the sectors equilibrate and hence no leak-in mechanism is observed. 
On the other hand, if portal coupling is very small then there would never 
be sufficient DM produced at early Universe that could reach observed 
relic abundance of DM after freeze-out. 

Analytically,  we can  obtain a lower limit on portal coupling by comparing 
observed relic density with the maximum relic abundance of DM in leak-in 
scenario. To do this, we first need to find the maximum  value attained by equilibrium yield,
\bea
\label{Yeq}
Y_\x^{\rm eq} = \dfrac{g_\x}{g_{*s} T^3} \dfrac{45}{2\pi^2}\left(\dfrac{m_\x \tp}{2 \pi}\right)^{3/2}e^{-m_\x/\tp}\,\,.\
\eea 
We maximize $Y_\x^{\rm eq}$ with respect 
to $\tp$ to obtain the dark sector temperature $\tp_{\rm max}$ at which the equilibrium yield is maximum. The value
of $\tp_{\rm max}$ is given by
\bea
\label{Tp_max}
\tp_{\rm max} = \dfrac{2m_\x\left(6-n\right)}{3n+30}\,\,.\
\eea 
One can notice that Eq.\,\ref{Tp_max} reduces to standard result of $\tp_{\rm max} = \frac{2m_\x}{5}$ \,\cite{Evans:2019vxr} 
as we choose $n=0$. Another point to note is that as $n$ increases,  $\tp_{\rm max}$ 
decreases. Using Eq.\,\ref{Tp_max}, the equilibrium yield at $\tp_{\rm max}$ is given by,
\bea
\label{Ymax}
{Y_\x^{\rm eq}}_{\rm max} &=& \dfrac{g_\x}{g_{*s}} \dfrac{45}{2\pi^2}\left(\dfrac{m_\x}{2 \pi}\right)^{3/2}
e^{-m_\x/\tp_{\rm max}}\dfrac{{\tp_{\rm max}}^{-(30+3n)/2(6-n)}}{K^{-24/(6-n)}}\,\,,\
\eea
where $K = \frac{10^4\sqrt{\eps}}{\left(1+\frac{n}{2}\right)^{1/4}}T_r^{n/8}$. 

Now, in case of fast expanding Universe, the DM can be produced via leak-in mechanism if ${Y_\x^{\rm eq}}_{\rm max} \geq Y_{\rm obs}$
where $Y_{\rm{\rm obs}} = 4.355 \times 10^{-10} \left(1 {\rm GeV}/m_\x\right)$. From this condition, one can derive an upper limit
on $\eps$ and it is given by
{\small\bea
	\label{eps-bound}
	\eps \gtrsim 2.616 \times 10^{-12} \left(3.561 \times 10^8\right)^{n/12}e^{n/8}\sqrt{\left(2+n\right)}\,\
	\left(\dfrac{g_\x}{g_{*s}}\right)^{\dfrac{n-6}{12}}
	\left(\dfrac{2 \left(6-n\right)}{3\left(10+n\right)}\right)^{\dfrac{10 + n}{8}} T_r^{-n/4} m_\x^{n/3}\,\,.\
	\eea}
Thus, Eq.\,\ref{eps-bound} tells us that unlike DM freeze-out in non-adiabatic 
scenario in standard cosmology, lower bound on $\eps$ is proportional to
$m_\x$. Thus, for leak-in mechanism, we need larger value of $\eps$ as we increase the mass of the DM.
The upper limit given in Eq.\,\ref{eps-bound} is known as ``Absolute coupling floor" below which
it is not possible to get correct relic density via leak-in mechanism.

\section{Numerical result for relic density}
\label{Numerical-result-for-relic-density}
In this section,  we discuss the evolution of DM comoving number density in fast expanding Universe where 
visible sector and dark sector are not in thermal equilibrium. The temperature evolution of DS is
different from the evolution of $T$ and we use the semi-analytic expression of $\xi$, given in Eq.\,\ref{Tprime}
to calculate $\tp$.
\subsection{Boltzmann equation}
\label{Boltzmann-equation}
Evolution of the DM number density depends on the production of DM from SM fields 
and annihilation of DM into $\zp$. Relevant Boltzmann equation for the evolution of total number density of 
DM is given by,
\bea
\label{BE}
\dfrac{dY_{\x_{\rm tot}}}{dx} &=&
\dfrac{s(x) h_{\rm eff} (x)}{H(x)x}\left[\dfrac{1}{2}\left[
\langle \sigma v\rangle ^{\tp}_{\x\bar{\x}\to \zp \zp}
\left(Y_{\x_{\rm tot}}^{\rm eq} (x,\xp)^2 - Y_{\x_{\rm tot}}^2\right)
\right] 
+
2 \sum_f \langle \sigma v\rangle^T_{{f}\bar{f}\to {\x}\bar{\x}}
Y_{f}^{\rm eq}(x)^2\right]\,,\nn\\
\eea 
where $Y_{\x_{\rm tot}}^{\rm eq} (x, \xp) = \dfrac{n_{\x_{\rm tot}}^{\rm eq}(\xp)}{s(x)}$
is the total equilibrium comoving number density of DM where, 
$n_{\x_{\rm tot}}^{\rm eq}(\xp) = n_\x^{\rm eq}(\xp) + n_{\bar{\x}}^{\rm eq}(\xp) $
is the total equilibrium DM number density at $\tp$. $Y_{f}^{\rm eq}(x) = \dfrac{n_f^{\rm eq}(x)}{s(x)}$ where, $n_f^{\rm eq}(x)$ 
is the equilibrium number density of SM fermion $f$ at SM temperature $T$, $h_{\rm eff} (x)= \left(1 - \dfrac{1}{3} \dfrac{d \ln g_{*s}(x)}{d \ln x}\right)$, and
$H(x)$ is the Hubble parameter, defined in Eq.\,\ref{eq:hubble}.

The first term inside the square bracket on the right hand side (RHS) of Eq.\,\ref{BE} denotes the contribution to the change 
in $Y_{\x_{\rm tot}}$ due to the annihilation of DM into $\zp$ and we call this term as ``annihilation term" in the rest of 
our discussion. $\langle \sigma v\rangle ^{\tp}_{\x\bar{\x}\to \zp \zp}$ 
is the thermally averaged cross section of $\x\bar{\x} \to \zp \zp$, 
calculated at $\tp$ and the $s$-wave approximated form of $\langle \sigma v\rangle ^{\tp}_{\x\bar{\x}\to \zp \zp}$ is given by
\bea
\label{xx-zz-themr-avg}
\langle \sigma v \rangle_{{\x}\bar{\x} \to \zp \zp}
&\simeq& \dfrac{4 \pi \alpha_X^2}{m_\x^2}
\dfrac{m_\x\left(m_\x^2 - m_{\zp}^2\right)^{3/2}}
{\left(2 m_\x^2 - m_{\zp}^2\right)^2}\,\,.
\eea
The second term inside the square bracket on the RHS of Eq.\,\ref{BE} denotes the production of DM from the SM bath and we call this term as ``freeze-in" term
in the rest of our discussion. Here, $\langle \sigma v\rangle^T_{{f}\bar{f}\to {\x}\bar{\x}}$ is the thermally 
averaged cross section of $f\bar{f} \to \x\bar{\x}$, calculated at temperature $T$. The cross sections of the relevant processes for the freeze-in are given by
\bea
\label{term-avg}
\sigma_{{f}\bar{f}\to {\x}\bar{\x}}
&=&
\dfrac{\kappa \alpha_X \eps^2}{3\hat{s}}\sqrt{\dfrac{\hat{s} - 4 m_\x^2}{\hat{s} - 4 m_f^2}}
\dfrac{(\hat{s} + 2 m_f^2)(\hat{s} + 2 m_\x^2)}{(\hat{s} - m_{\zp}^2)^2} \,\,,
\eea 
where $m_f$ is the mass of the SM fermion $f$, $\kappa = 1$ for $f=\mu,\,\tau$ and $\kappa=1/2$ for $f=\nu_\mu,\,\nu_\tau$.

In the following section, we discuss the effect of annihilation and freeze-in terms on the DM relic density in the
fast expanding Universe. The numerical results related to the DM relic density has also been discussed in the next section.

\subsection{Outcome of numerical analysis}
In this section, we present our numerical analysis by solving Boltzmann equation given in Eq.\,\ref{BE}
in the presence of fast expanding Universe. In our analysis, we use the semi-analytic expression of $\xi(T)$,
given in Eq.\,\ref{Tprime}, to calculate $\tp$. Here, we have considered $T_r=20\rm MeV$ as discussed earlier and $m_{\zp} = 100\rm MeV$. Let
us note that the choice of $m_{\zp}$ is motivated by the explanation of muon $(g-2)$ anomaly. However, as far as the relic density analysis is concerned,
it is independent on the choice of $m_{\zp}$ for $m_\x/m_{\zp} \gtrsim 10$.

 \begin{figure}[h!]
        \centering
        \includegraphics[height = 7cm, width = 8cm]{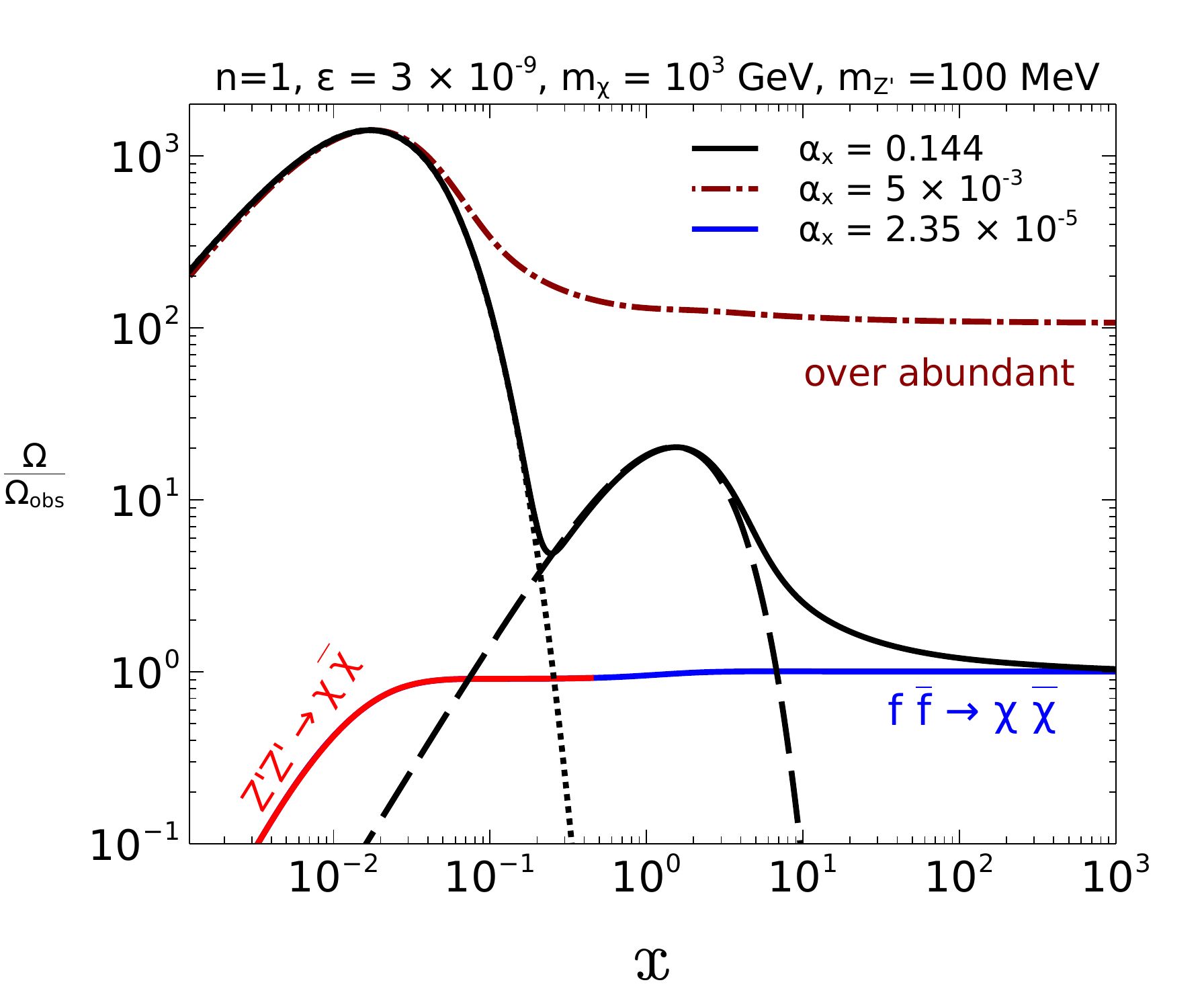}
        \includegraphics[height = 7.2cm, width = 8cm]{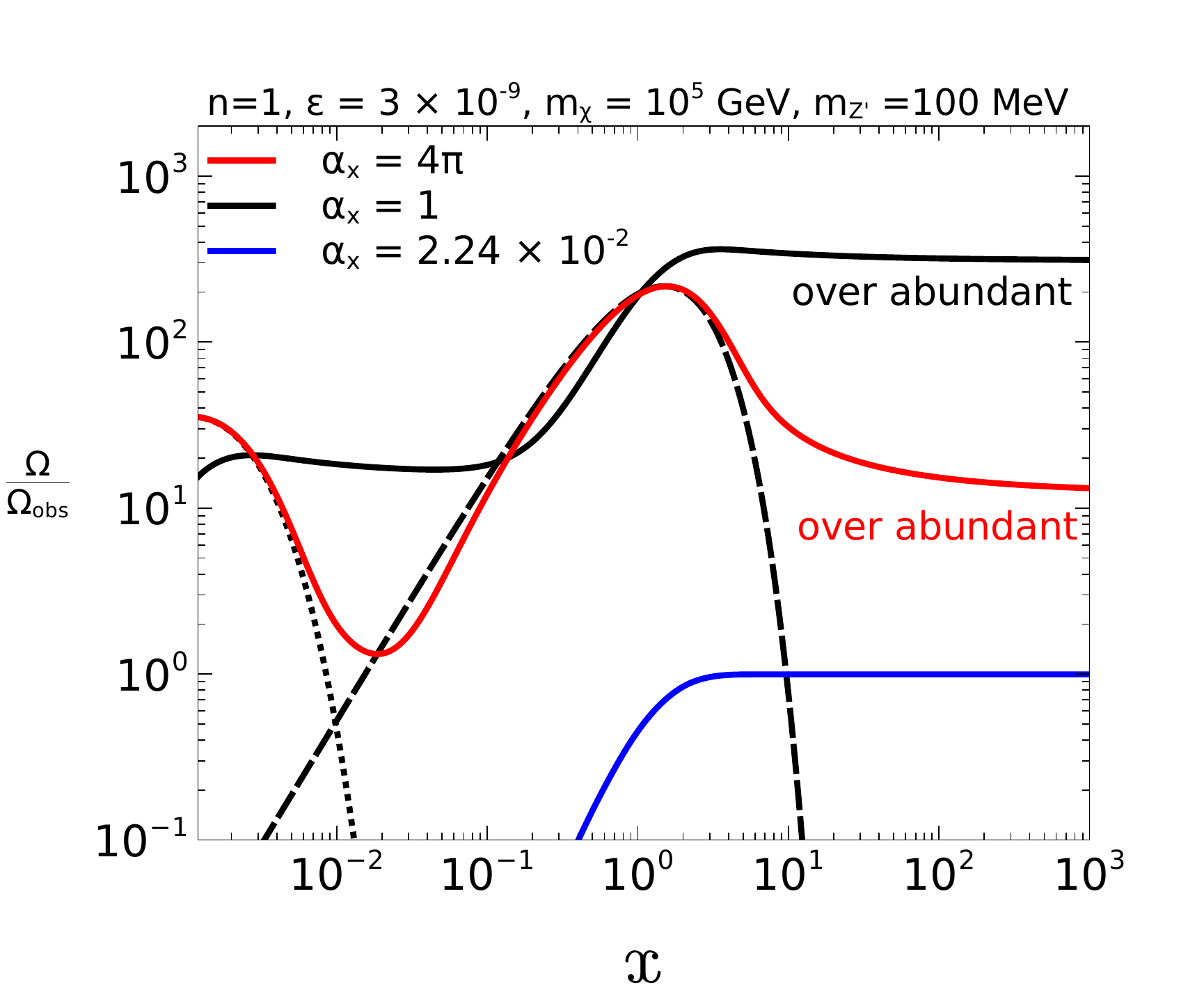}
        \caption{Evolution of relative abundance of DM as a function of 
                 $x$ for $n=1$, $\eps = 3 \times 10^{-9}$, and $m_{\zp}=100\rm MeV$.
                \textit{\textbf{Left panel}}: Here we consider $m_\x=10^3\rm GeV$ and for 
$\alpha_X = 2.35 \times 10^{-5}$ (red and blue solid line) and $\alpha_X = 0.144$ (black solid line), DM is
produced via freeze-in and reannihilation mechanisms respectively. For $\alpha_X = 5 \times 10^{-3}$ 
(brown dot-dashed line), final abundance of DM is determined by the leak-in mechanism but it overcloses the Universe.
\textit{\textbf{Right panel}}: Here, we consider $m_\x = 10^5\rm GeV$. 
                For $\alpha_X = 2.24 \times 10^{-2}$ DM is produced by freeze-in mechanism and it is shown by 
                blue solid line. For $\alpha_X = 1$, (black solid line) and $4 \pi$ (red solid line), the amount of produced DM 
                overcloses the Universe. In both panels, black dotted and black dashed lines denote the evolution of 
                $\Omega h^2_{\rm eq}/\Omega h^2_{\rm obs}$ and $\Omega h^2_{\rm QSE}/\Omega h^2_{\rm obs}$ respectively.}
        \label{BE-plot_n1}
\end{figure}

In Fig.\,\ref{BE-plot_n1}, we have shown the variation of the relative abundance of DM (defined as $\Omega h^2/\Omega h^2_{\rm obs}$, 
where $\Omega h^2 = 2.755 \times 10^8 (m_\x/1\rm GeV) Y_{\x_{\rm tot}}$ and $\Omega h^2_{\rm obs} = 0.12$) as a function of $x$
for $n=1$, $\eps=3\times 10^{-9}$ and two different values of $m_\x$. In the left panel of Fig.\,\ref{BE-plot_n1}, we have considered 
$m_\x = 10^3 \rm GeV$. Here, one can see that for three different values of $\alpha_X$, DM production mechanisms are different. 
When $\alpha_X$ is very small, DS is not internally thermalized and DM can be produced by both $\zp \zp \rightarrow \x \bar{\x}$ 
and $f \bar{f} \rightarrow \x \bar{\x}$ processes. At large temperature i.e. when $(\tp \gtrsim m_\x)$,
DM production is dominated by $\zp \zp \rightarrow \x \bar{\x}$ process and it is shown by the red solid line.
Now, due to the presence of the freeze-in term in the Boltzmann equation, DM can also be produced from the annihilation of SM fermions
and this production will stop at $T\sim m_\x$.  The solid blue line indicates the production of DM from SM annihilation.
Therefore, for $\alpha_X = 2.35 \times 10^{-5}$, DM is produced via freeze-in mechanism.
Now, if we increase the value of $\alpha_X$, one of the possible production mechanisms
leak-in is observed, which is shown by brown dot-dashed line. One can notice that DM produced
through leak-in mechanism is over abundant. Now, for $\alpha_X =0.144$, variation of the 
relative abundance of DM is shown by black solid line. As we can see, initially the relative abundance follows its equilibrium value
i.e. the evolution of $\Omega h^2_{\rm eq}/\Omega h^2_{\rm obs}$ where $\Omega h^2_{\rm eq} = 2.755 \times 10^8 (m_\x/1\rm GeV) Y^{\rm eq}_{\x_{\rm tot}}$
and it is depicted by black dotted line. When $\tp < m_\x$, number density of DM started to fall off 
exponentially but due to the presence of freeze-in term $Y_{\x_{\rm tot}}$ starts to increase and at this time it follows a quasi static equilibrium.
Quasi static equilibrium is established when DM annihilation rate due to annihilation term and DM production rate due 
to freeze-in term are equal\,\cite{Cheung:2010gj}. DM comoving number density in this quasi static equilibrium is denoted by 
$Y_{\rm{QSE}}$ and it is independent of the expansion parameter $n$. The analytic form of $Y_{\rm QSE}$ is given by
\bea
\label{yqse}
Y_{\rm {QSE}} &=& \dfrac{ \langle \sigma v\rangle^T_{f\bar{f}\to \x\bar{\x}}n_{f}^{\rm eq}(T)^2}
{\langle \sigma v\rangle ^{\tp}_{\x\bar{\x}\to \zp \zp}s(T)^2}\,\,.\
\eea
In this figure we show the variation of the quantity $\Omega h^2_{\rm QSE}/\Omega h^2_{\rm obs}$ by the dashed black line and 
$\Omega h^2_{\rm QSE} = 2.755 \times 10^8 (m_\x/1{\rm GeV}) Y_{\rm QSE}$.
Here, one can see that nearly at $x\simeq 10$, DM departs 
from quasi static equilibrium and freezes out. This phenomena is 
known as reannihilation. In the right panel of Fig.\,\ref{BE-plot_n1}, we have shown that for $m_\x = 10^5 \rm GeV$, only 
possible way to produce correct relic abundance of DM is freeze-in and it is shown by solid blue line.
For $\alpha_X = 1$ (black solid line) and $\alpha_X = 4\pi$ (red solid line), 
excess amount of DM produced from the SM bath cannot annihilate sufficiently into $\zp$ to give correct 
relic density. Thus for $n=1$, $m_\x=10^3\rm GeV$ and $10^5\rm GeV$ reannihilation and freeze-in are the 
only possible way to produce observed relic density. Let us note that, to satisfy the relic density constraint, 
we need larger values of $\alpha_X$ in comparison to the standard freeze-in scenario 
\cite{Hall:2009bx}. 

\begin{figure}[h!]
        \centering
        \includegraphics[height = 7cm, width = 8cm]{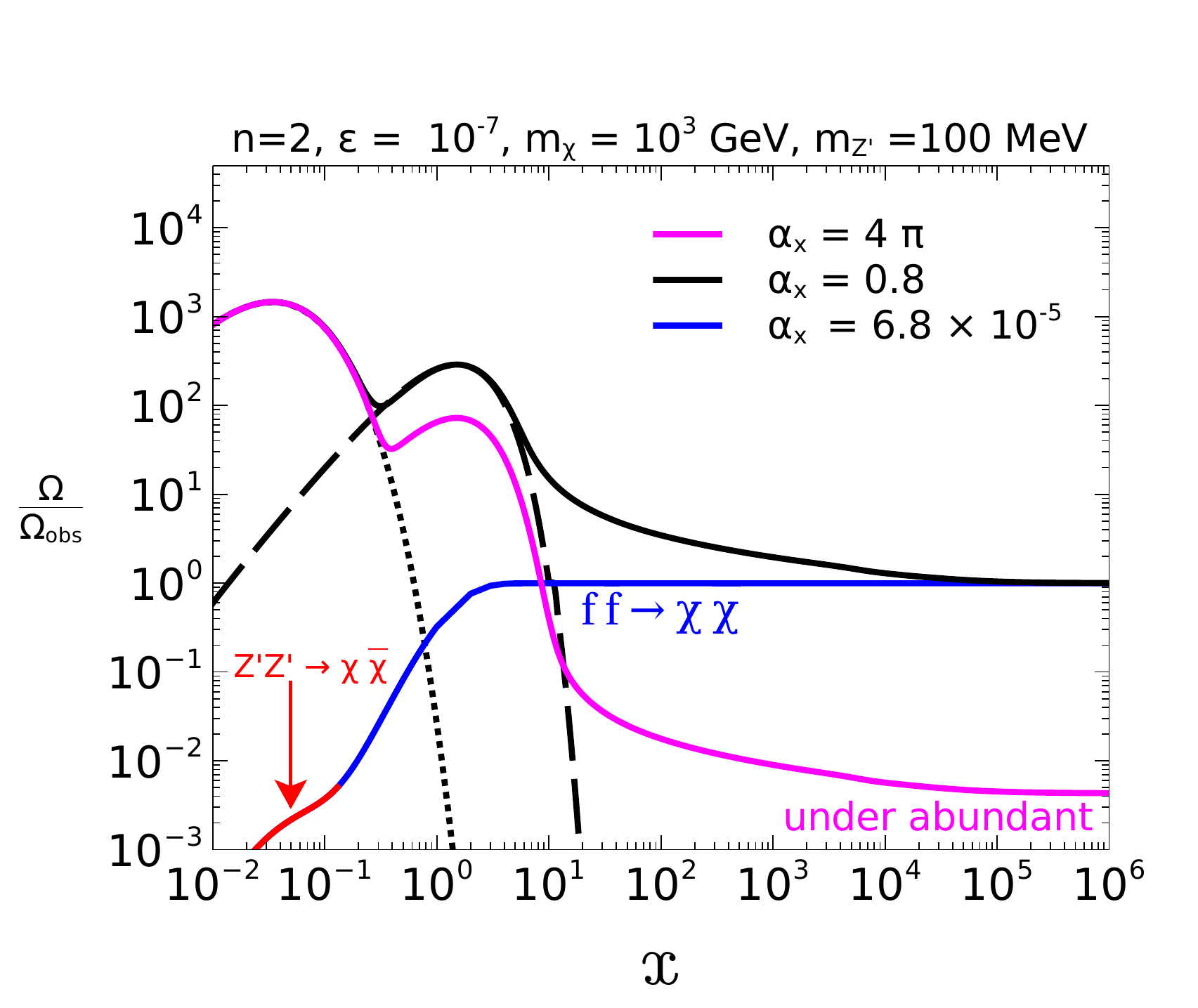}
        \includegraphics[height = 7cm, width = 8cm]{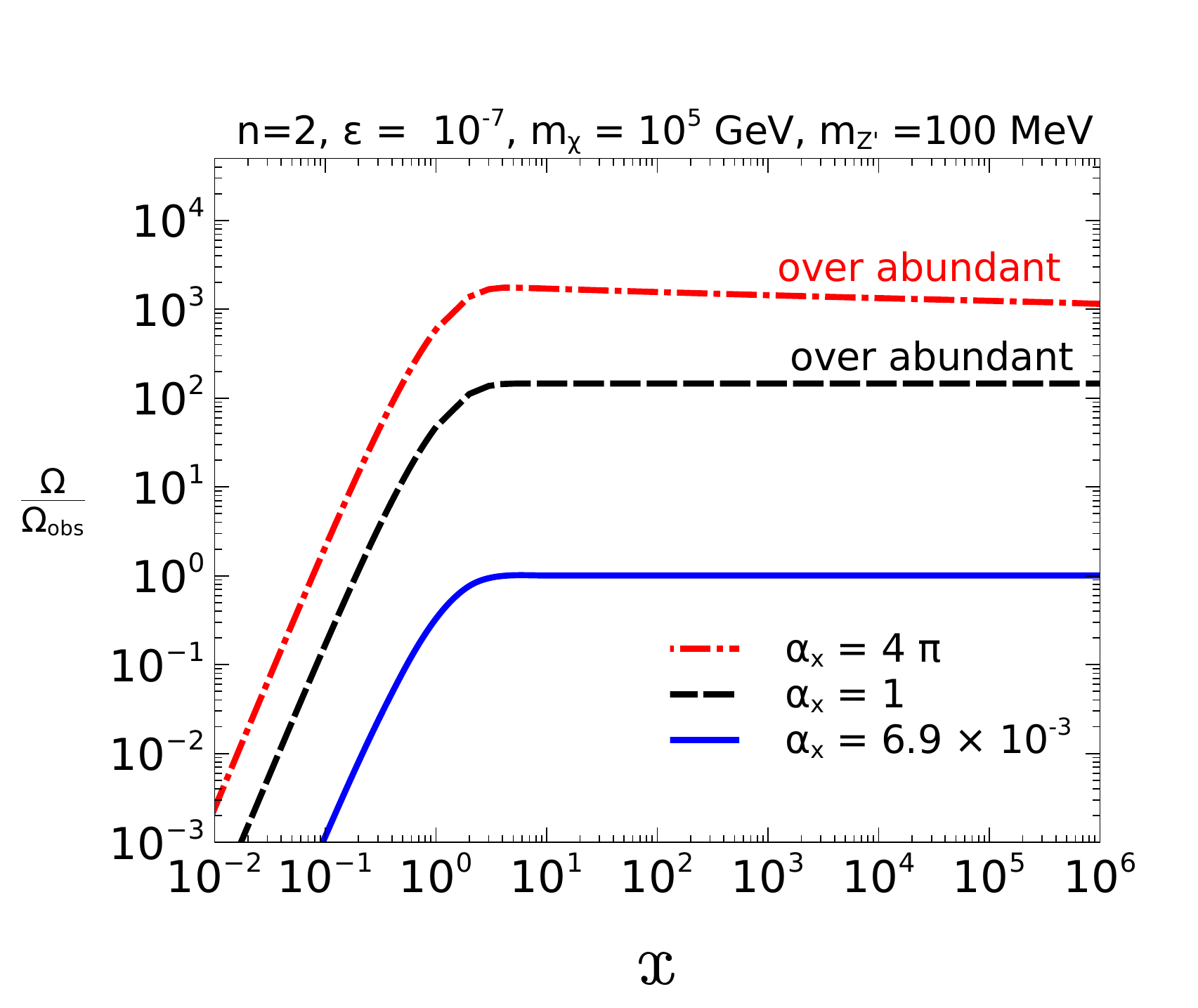}
        \caption{
                \textit{\textbf{Left panel}}: Evolution of relative abundance of DM for $m_\x = 10^3 \rm GeV$.
                For $\alpha_X = 6.8 \times 10^{-5}$ and $0.8$ DM production
                mechanisms are freeze-in (blue solid line) and reannhilation (black solid line) respectively.
                Red line over blue line indicates DM production from $\zp$.
                For $\alpha_X = 4 \pi$, DM is underabundant and it is shown with solid magenta line. The evolution of $\Omega h^2_{\rm eq}/\Omega h^2_{\rm obs}$
and $\Omega h^2_{\rm QSE}/\Omega h^2_{\rm obs}$ are denoted by black dotted and black dashed lines respectively
                \textit{\textbf{Right panel}}: Evolution of DM relative abundance for $m_\x = 10^5 \rm GeV$. For
                $\alpha_X = 6.9 \times 10^{-3}, 1,\,\text{and }4 \pi$, DM production mechanisms are freeze-in
                (blue solid line), late-time annihilation (black dashed line), and late-time annihilation
                (red dot-dashed line) respectively. Here we choose $n=2$, $\eps = 10^{-7}$, and $m_{\zp} = 100\rm MeV$ in both the figures. 
       }
        \label{BE-plot_n2}
\end{figure}

For $n=2$, we show the evolution of relative abundance of DM for $\eps = 10^{-7}$ in Fig.\,\ref{BE-plot_n2}.
In the left panel of Fig.\,\ref{BE-plot_n2}, we have considered $m_\x = 10^{3}\rm GeV$. Here, we can
see that for $\alpha_X = 6.8 \times 10^{-5}$, DM is produced via freeze-in and it is shown by solid
blue line. The red line over blue line indicates the DM production from $\zp$ annihilation in the early Universe. For $\alpha_X=0.8$,
DS can internally thermalize and one of the possible production mechanisms is reannihilation, shown by
black solid line. Evolution of $\Omega h^2_{\rm QSE}/\Omega h^2_{\rm obs}$ and $\Omega h^2_{\rm eq}/\Omega h^2_{\rm obs}$
are shown by the black dashed and black dotted lines respectively. DM annihilation into $\zp$ 
increases if we increase $\alpha_X$ and the magenta line represents the evolution of relative abundance of DM for
$\alpha_X=4\pi$. Thus, possible ways of DM production in this case are freeze-in and reannihilation. In the right panel of Fig.\,\ref{BE-plot_n2},
we have considered $m_\x = 10^5\rm GeV$. One can see from this figure that initially the DM does not follow the equilibrium comoving number density.
This is because the annihilation cross section $\x \bar{\x}\to \zp \zp$ decreases as we increase $m_\x$ and as a result the DM annihilation
cross section is not sufficient to thermalise the DM. Here, we have shown that for $\alpha_X = 6.9 \times 10^{-3}$ and $1$, relic abundance of DM can be obtained 
via freeze-in mechanism and they are denoted by solid blue and dashed black lines respectively. For $\alpha_X = 4\pi$, excess amount of DM is produced from the SM bath
and it is represented by red dot-dashed line. Now, due to large value of $\alpha_X$, excess DM produced from the SM bath, 
starts to annihilate into $\zp$. We call this phenomenon as ``late-time annihilation". However, within the perturbative limit of $\alpha_X$, 
annihilation cross section of $\x\bar{\x}\to \zp\zp$ is not sufficiently large to give correct relic abundance of DM. Here, one can
see that for $\alpha_X=1$ and $4\pi$, DM is overabundant. 

\begin{figure}[hbt!]
        \centering
        \includegraphics[height = 7cm, width = 8cm]{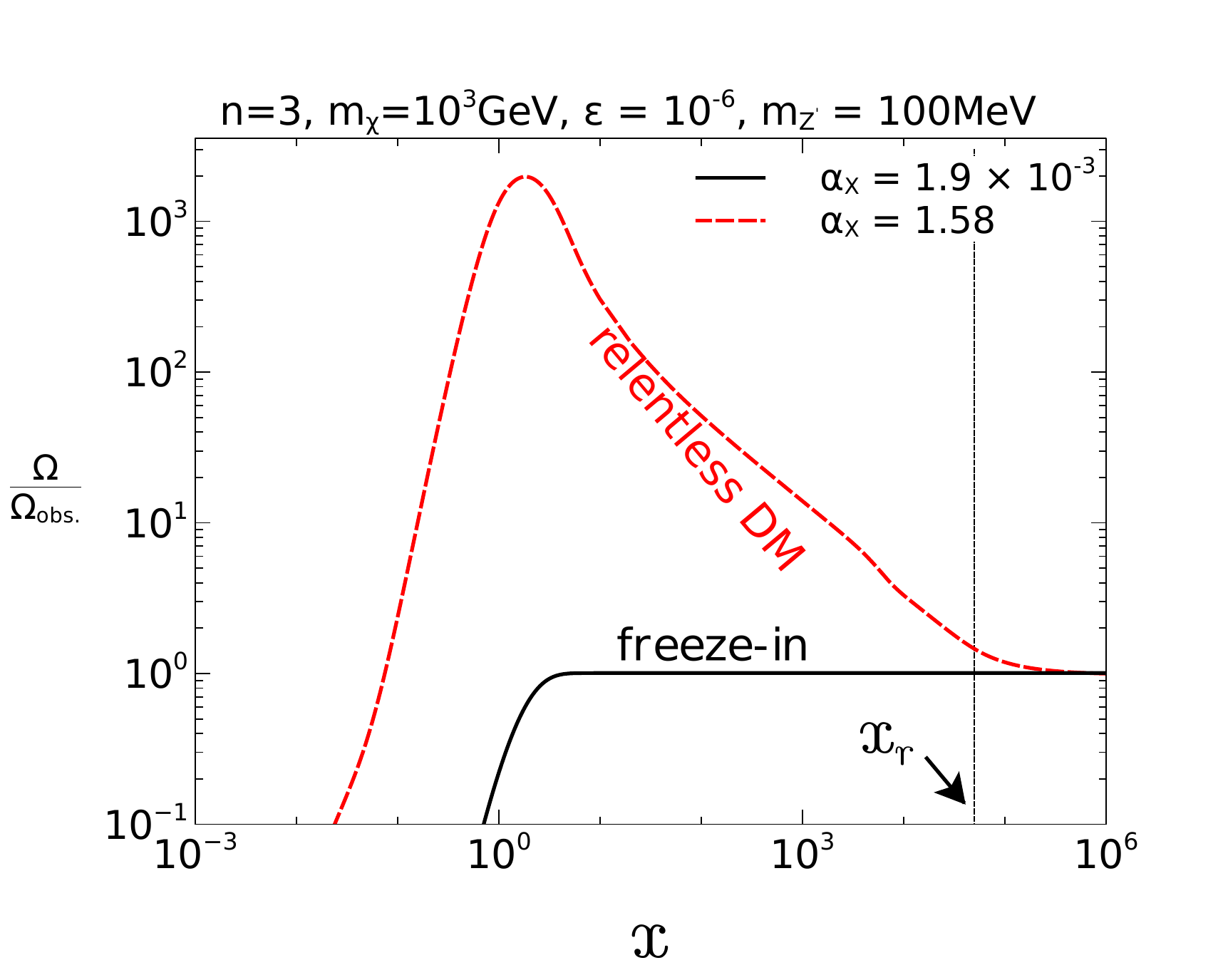}
        \includegraphics[height = 7cm, width = 8cm]{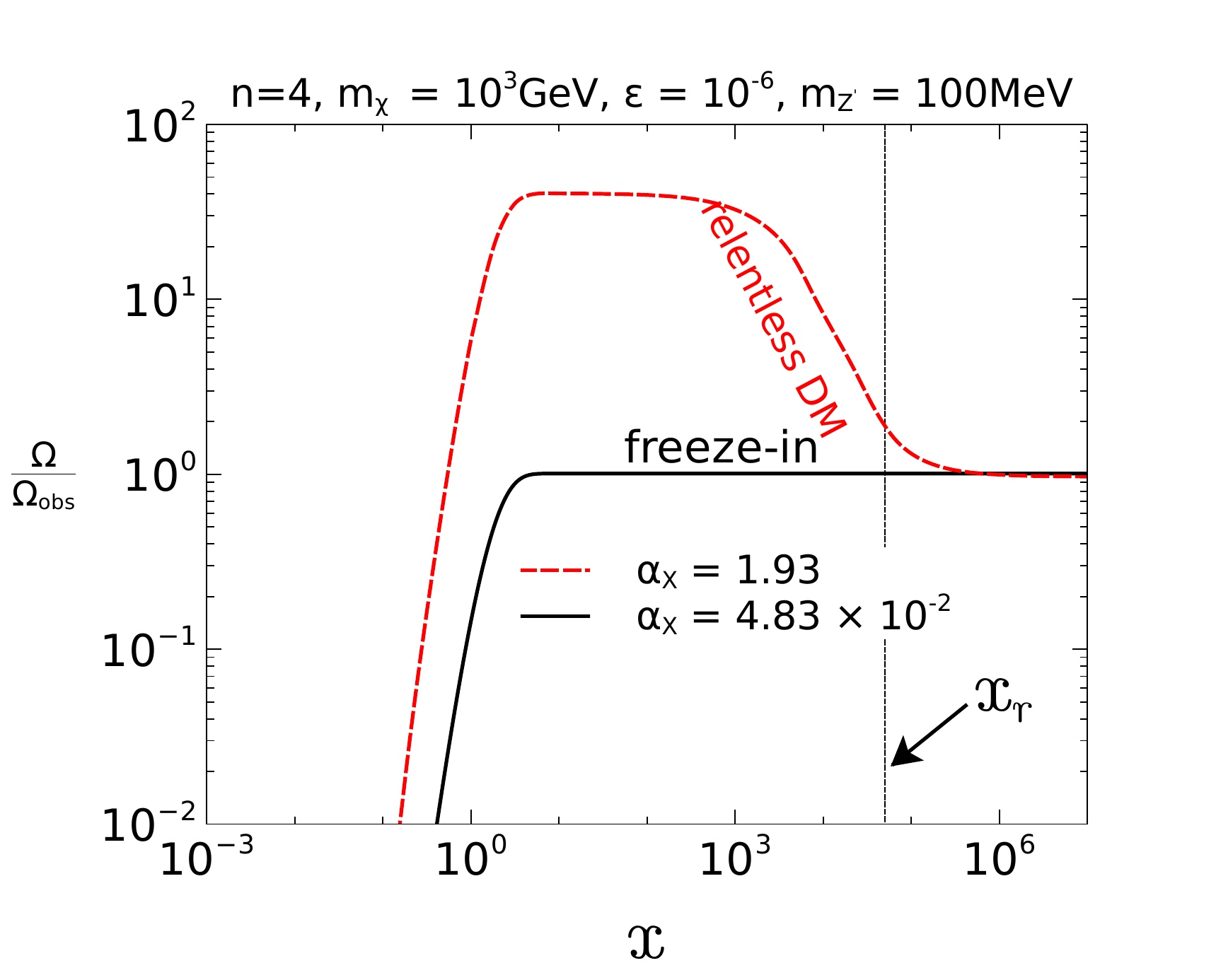}
        \caption{Relative abundance of DM as a function of $x$ for $m_\x=10^3\rm GeV$, $\eps=10^{-6}$, and $m_{\zp}=100\rm MeV$.
        \textbf{\textit{Left panel}}: We consider $n=3$. For $\alpha_X = 1.9 \times 10^{-3}$ (solid black line) and $\alpha_X = 1.58$ (red dashed line) 
        DM is produced via freeze-in and late-time annihilation mechanisms respectively.
        \textbf{\textit{Right panel}}: DM is produced via freeze-in and late-time annihilation mechanisms for $\alpha_X=4.83 \times 10^{-2}$ (black solid line) and
         $1.93$ (red dashed line) respectively. Here we have considered the expansion parameter $n=4$. In both panels $x_r=m_\x/T_r$. 
        }
        \label{BE-plot_n3-4}
\end{figure}

As $n$ increases,  parameter space for internally thermalized DS reduces. 
As a result, the possible ways of producing DM are freeze-in and late-time annihilation. As we can see in the left panel of 
Fig.\,\ref{BE-plot_n3-4}, for $m_\x = 10^3  \rm{GeV}$, one of the possible mechanisms here is freeze-in when $\alpha_X=1.9 \times 10^{-3}$, 
and it is shown in black solid line. The red dashed line depicts a new possible way of producing DM for $n>2$. 
Here we see that for $\alpha_X = 1.58$, the dark sector is not internally thermalized and DM is produced from SM bath via freeze-in mechanism.
Due to the large value of $\alpha_X$, the excess amount of DM produced via freeze-in can annihilate into $\zp$  and finally DM freezes out and 
the final abundance of DM is set by the late-time annihilation. Furthermore, due to the presence of the fast expanding component $\phi$ in the
Hubble parameter, DM keeps annihilating and as a result the freeze-out process is prolonged. This is known as ``relentless DM" \cite{DEramo:2017gpl}.
In right panel of Fig.\,\ref{BE-plot_n3-4} similar features are shown for $n=4$. Here, we can see a plateau region for large
$\alpha_X$, this is because freeze-in mechanism stops when $x \simeq 1$
and at this time annihilation rate of DM is much less than $H$. However, it is expected that the annihilation of DM will start again due to
the presence of the fast expanding component in the Hubble parameter. We have numerically checked that
$\Gamma_{\x \bar{\x} \to \zp \zp} \simeq  H$ near $x \simeq 10^3$ and 
DM starts to annihilate into $\zp$ and final relic density of DM is determined by the late time annihilation process.
\begin{figure}[h!]
        \centering
        \includegraphics[height = 10cm, width = 12cm]{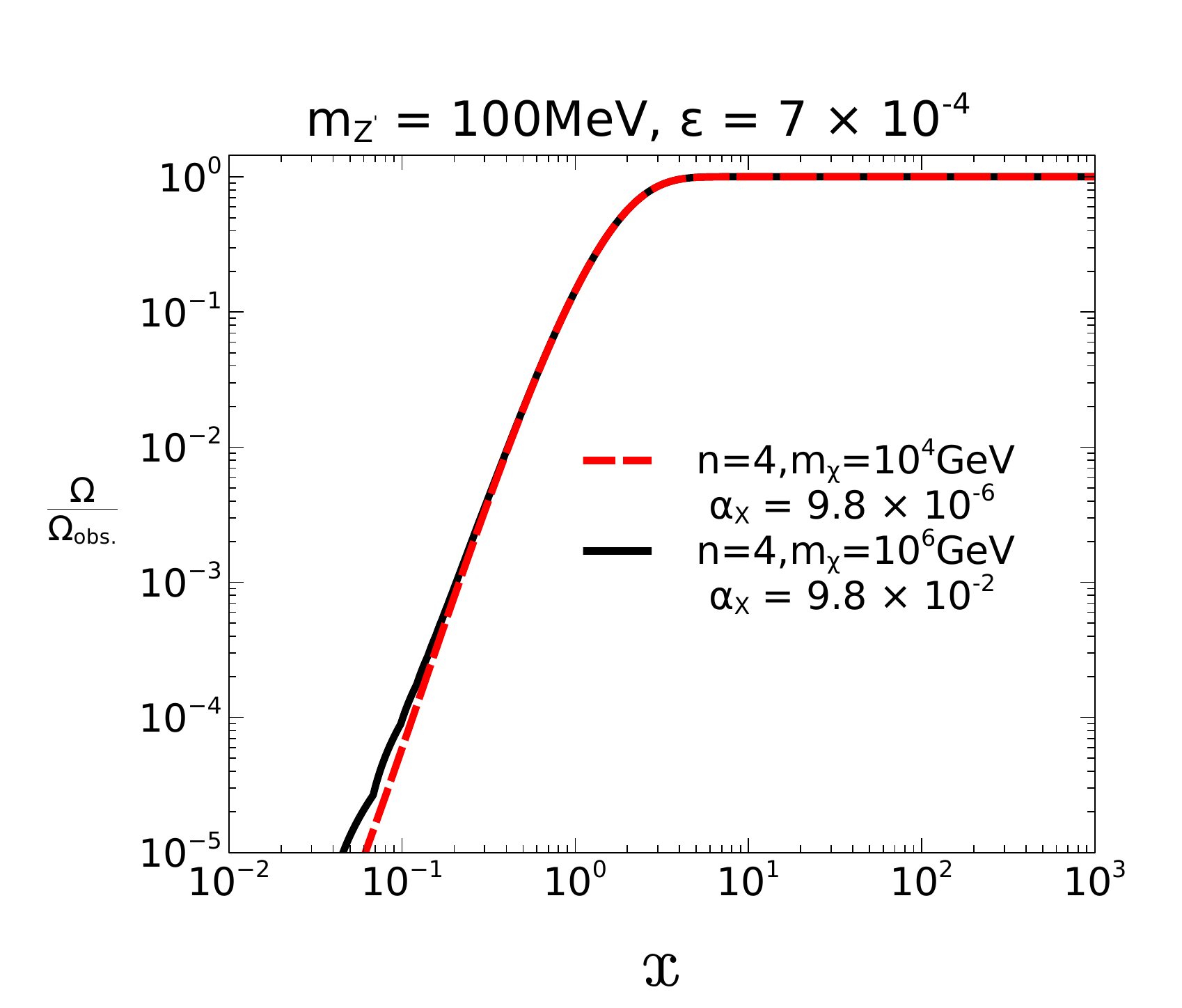}
        \caption{ Evolution of relative abundance of DM as a function of $x$ for $\eps=7 \times 10^{-4}$ and $m_{\zp}=100\rm MeV$, and $n=4$
        for two different set of values of $(m_\x, \,\alpha_X)$. The evolution for $(10^4{\rm GeV}, \,9.8 \times 10^{-6})$ and $(10^6{\rm GeV}, \,9.8 \times 10^{-2})$
        are denoted by red dashed line and solid black line respectively.
        }
        \label{relic-g-2-fig}
\end{figure}
\begin{figure}[h!]
        \centering
       \includegraphics[height = 10cm, width = 12cm]{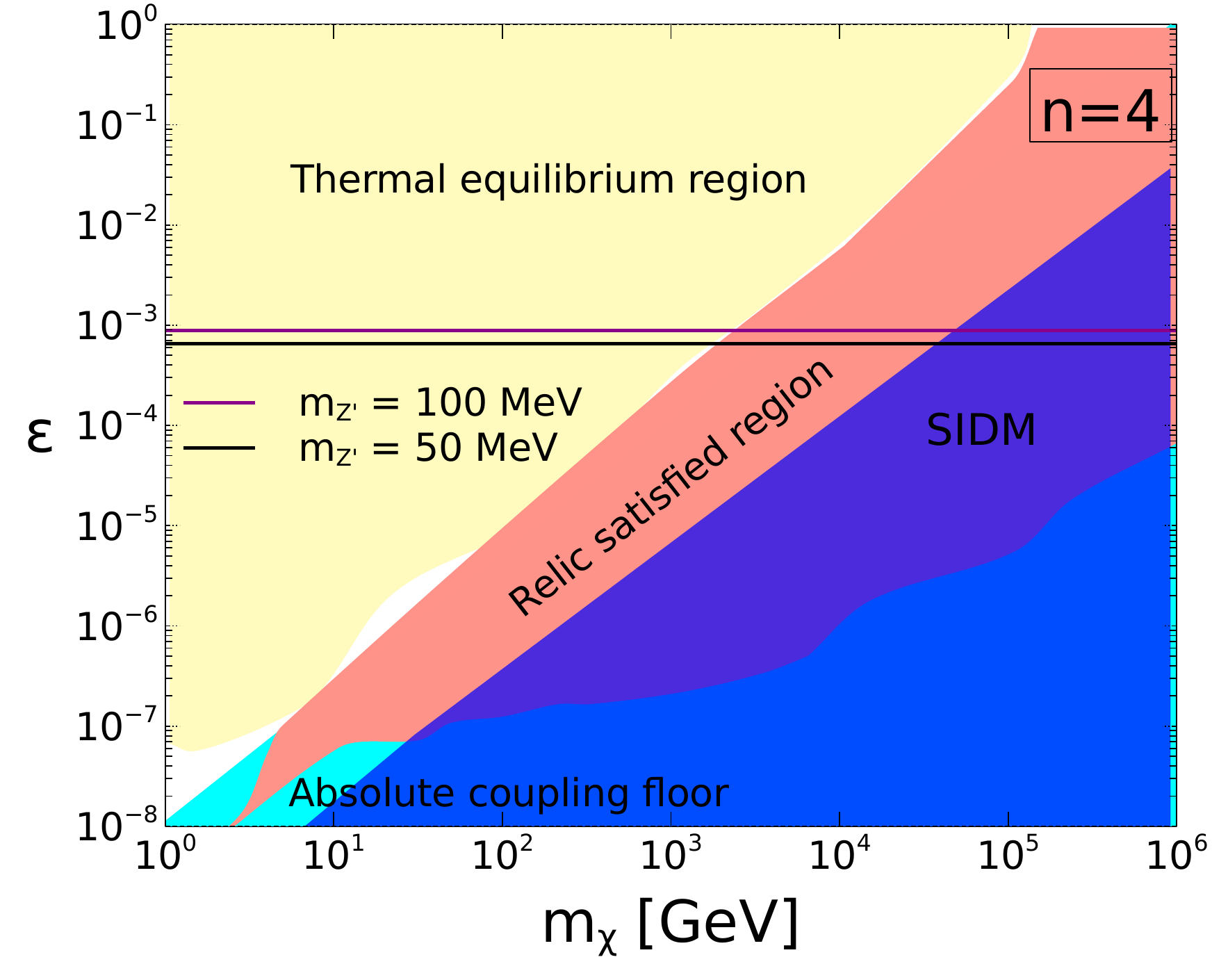}
        \caption{Constraints in the $m_\x-\eps$ plane. In the yellow region, dark and visible sectors are in thermal equilibrium. 
The relic density satisfied region is depicted by the light red region. Cyan region denotes the absolute coupling floor for the leak-in scenario.
DM self interaction constraint is shown by the blue region. The horizontal solid lines denote the required value of $\eps$
to satisfy the muon $(g-2)$ anomaly within $2\sigma$ limit for $m_{\zp}= 100\rm MeV$ (violet line) and $m_{\zp}= 50\rm MeV$ (black line). In this
figure, we have considered the value of the expansion parameter $n=4$.}
\label{reanh-plot}
\end{figure}

In section \ref{wimp-next-door}, we mentioned a certain region of parameter space for a thermally decoupled dark sector in which it is possible to explain 
muon $(g-2)$ anomaly. Motivated by this result, in Fig.\,\ref{relic-g-2-fig}, we show the evolution of relative DM abundance
for different values of $m_\x$ and $\alpha_X$ and the choices of $m_{\zp}$ and $\eps$ are such that it can enhance
the theoretical prediction of muon $(g-2)$. One can see that in this figure, the DM is produced via freeze-in mechanism.
This is  because for large values of $n$, internal thermalization of dark sector is not possible within the perturbative limit of
$\alpha_X$. An important point to note is that in standard cosmology, DM relic density produced via freeze-in mechanism, is independent of 
DM mass and it only depends on coupling strength \cite{Hall:2009bx}. However, for fast expansion, the relic density depends on 
$m_\x$, $\eps$, and $\alpha_X$. Therefore, for fixed $\eps$ we need larger value of $\alpha_X$ as we increase $m_\x$. The calculations
related to freeze-in in fast expanding Universe is discussed in appendix \ref{Appx-A}.

In Fig.\,\ref{reanh-plot} we have shown DM relic density satisfied region and different constraints
in case of $n=4$ in $m_\x -\eps$ plane. As we have discussed earlier,  DM production mechanism for
$n=4$ is freeze-in and late time annihilation. Relic satisfied region is shown by the light red colored region.
Here, one can see that the required value of $\eps$ increases as we increase DM mass and this can be
easily understood from Eq.\,\ref{final-relic-freezein}. In the yellow shaded region of the figure,
dark and visible sectors are in thermal equilibrium. The absolute coupling floor for leak-in scenario is shown
by the cyan region. The horizontal violet and black solid lines denote the required values of $\eps$
for $m_{\zp}=100\rm MeV$ and $50\rm MeV$ to satisfy the muon $(g-2)$ anomaly respectively.
We have also studied the constraint from self interaction of DM which is discussed later
and the constrained region in $m_\x-\eps$ plane is shown by the blue shaded region of Fig.\,\ref{reanh-plot}.
\section{Relevant constraints on model}
\label{constraints-on-lmt}
\subsection{Self interacting dark matter (SIDM)}

DM self interaction cross-section per unit DM mass can be constrained from bullet cluster
observation and the upper limit of $\sigma^T/m_\x$ is $1.25\,\rm cm^2 g^{-1}$ \cite{Randall:2008ppe} where $\sigma^T$ is the momentum
transfer cross section of DM elastic scattering processes. In our model,  DM elastic scattering processes are mediated by dark vector boson $\zp$.
We have considered $m_{\zp} = 100 \, \rm MeV$ and choice of $m_{\zp}$ is motivated by the explanation of muon $(g-2)$ anomaly.
We calculate total $\sigma^T$ and it is defined as
\bea
\sigma^T = \dfrac{1}{4}\left[\sigma^T_{\x \bar{\x} \to \x \bar{\x}} + \sigma^T_{\x \x \to \x \x}+\sigma^T_{\bar{\x} \bar{\x} \to \bar{\x} \bar{\x}}\right]\,.
\eea
The blue region of Fig.\,\ref{reanh-plot} depicts the excluded region of parameter space from DM self interaction. In deriving
the SIDM bound, we have expressed $\alpha_X$ in terms of $m_\x$ and $\eps$ using Eq.\,\ref{final-relic-freezein} and the relic density constraint
$\Omega h^2_{\rm obs} = 0.12$.

\subsection{Muon anomalous magnetic moment}
\label{muon-g-2}
In our model, the dark vector boson $\zp$ can contribute to the anomalous magnetic moment of muon if its coupling strength 
with muon is $\sim 10^{-3}$ and $m_{\zp}\sim 100\rm MeV$. As we discuss in the earlier sections, it is possible to satisfy
the relic density constraint for $\eps \sim 10^{-3}$ if we consider fast expanding Universe. Thus in our framework, one can have simultaneous
explanations of DM relic density and muon $(g-2)$.

The contribution of $\zp$ to $\Delta a_\mu$ is given by \cite{Baek:2001kca,Ma:2001md,Escudero:2019gzq, Banerjee:2020zvi}
\bea
\Delta a_\mu &=& \dfrac{\eps^2}{4 \pi^2} \int_{0}^{1} \dfrac{m_\mu^2 z\left(1-z\right)^2}{m_\mu^2\left(1-z\right)^2 + m_{\zp}^2 z} dz\,\,.\
\label{func_anomaly}
\eea

\begin{figure}
	\centering
	\includegraphics[height = 10cm, width = 12cm]{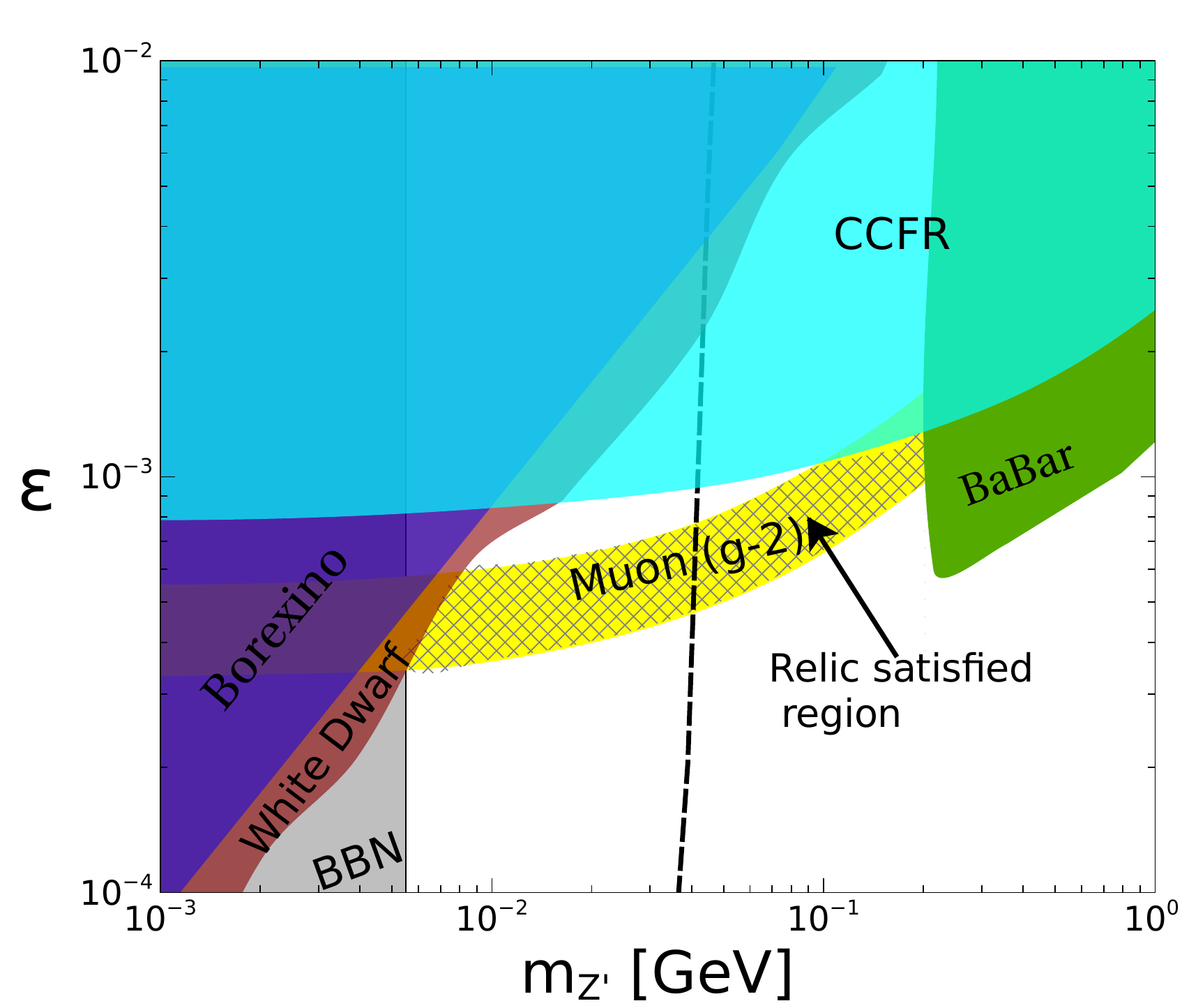}
	\caption{Constraints in $m_{\zp}-\eps$ plane. The cyan region denotes the constraint from the neutrino
trident production at CCFR \cite{CCFR:1991lpl,Altmannshofer:2014pba}. BBN Constraint is shown by the gray colored region.
Brown region depicts the constraint from white dwarf cooling. The observed deviation from the SM prediction of muon $(g-2)$
can be explained in the yellow region within $2\sigma$ limit. In the black hatched region, depending on the expansion parameter $n$
one can satisfy the relic density constraint. The left side of the black dashed line represents the region where 
$\zp$ is in thermal equilibrium neutrino bath. The constraints from Borexino \cite{Bellini:2011rx} and BABAR \cite{BaBar:2016sci} experiments 
discussed in \cite{Gninenko:2020xys}, are shown by violet and green regions respectively \cite{Gninenko:2020xys}. }
\label{constraint}
\end{figure}

The present value of $\Delta a_\mu = (251 \pm 59)\times 10^{-11}$ \cite{Muong-2:2021ojo} and the yellow region of Fig.\,\ref{constraint} depicts the muon $(g-2)$
satisfied region within $2\sigma$ limit in $m_{\zp}-\eps$ plane.

\subsection{White dwarf cooling}
Several astrophysical observations constrain $L_\mu - L_\tau$ portal. For example, SN1987A
constrains very low portal coupling \cite{Fradette:2014sza,Chang:2016ntp,Escudero:2019gzq}
whereas presence of $\zp$ inside stellar core affects the rate of stellar cooling 
\cite{An:2013yfc,Hardy:2016kme} and the parameter space of very low mass $\zp$ is severely constrained from the stellar cooling constraint.
For the mass range of $\zp$ in which we are interested in, most relevant constraint comes from white dwarf (WD) cooling.

The WD cooling constraint in our scenario can be derived from the following effective electron-neutrino interaction Lagrangian
\bea
\mathcal{L}^{\rm NP}_{\nu e} \supset -C_{\rm WD} \left(\bar{\nu}_l \gamma_\mu P_L \nu_l\right)(\bar{e} \gamma^\mu e)\,\,,
\label{eq:WD_cooling}
\eea
where
\bea
C_{\rm WD} =  \frac{\eps}{m_{\zp}^2} \sqrt{4 \pi \alpha_{\rm {em}}}\,\Pi(0)\,\,.
\eea

The $\gamma-\zp$ kinetic mixing at one loop level $\Pi (q^2)$ is given by \cite{Araki:2017wyg, Banerjee:2018mnw}
\bea
\label{mixing-loop}
i \Pi (q^2) = - \frac{i e \epsilon}{2 \pi^2}
\int_0^1 x (1-x) \ln\left(\dfrac{m_\mu^2 -q^2 x(1-x)}
{m_\tau^2 - q^2 x(1-x)} \right) dx\,\,\,.
\eea
In the low momentum transfer limit i.e. $q^2 \to 0$, Eq.\,\ref{mixing-loop} takes the following form
\bea
i \Pi (0) = - \frac{i e \epsilon}{12 \pi^2} 
\ln\left(\dfrac{m_\mu^2}{m_\tau^2}\right)\,\,.
\eea

From the rate of cooling of WD, the Wilson coefficient $C_{\rm WD}$ of neutrino-electron interaction is constrained as \cite{Dreiner:2013tja,Arguelles:2022xxa}
\bea
\label{WD}
\dfrac{1.12 \times 10^{-5}}{\rm {GeV}^2} < C_{\rm {WD}}
< \dfrac{4.5 \times 10^{-3}}{\rm {GeV}^2} \,\,.
\eea
We have shown the constraint coming from WD cooling by brown colored region in Fig.\,\ref{constraint}.
\subsection{BBN constraint}
Phenomenology of early Universe near $1 \rm {MeV}$ is highly constrained from BBN predictions \cite{Mukhanov:2003xs,Mangano:2011ar}.
Due to presence of light $\zp$, it is possible that $\zp$ can thermalise with the SM neutrinos via $\zp \leftrightarrow \bar{\nu}_i \nu_i$ ($i=\mu, \,\tau$) 
and contribute to the effective number of neutrino species ($N_\nu$) at the time of BBN. 

To check the thermalisation of $\zp$ at the time of BBN, we have compared the reaction rate of $\bar{\nu}_i \nu_i \leftrightarrow \zp$ with the Hubble parameter
at the time of BBN ({\it i.e.} at $T=1\rm MeV$) and found that the $\zp$ is in thermal equilibrium with the neutrino bath for $10^{-4}\leq \eps\leq 10^{-2}$ and 
${\rm 1MeV} \leq m_{\zp} \lesssim 40 \rm MeV $. This region is outlined by the black dashed line in Fig.\,\ref{constraint}. In this region of parameter
space, depending on the mass of $\zp$, it may give additional contribution to the effective number of neutrino species
($\Delta N_\nu \equiv N_\nu-3$) whose upper limit is 0.168 at 95\% C.L \cite{Fields:2019pfx}.
We have calculated the constraint on the $\zp$ mass from $\Delta N_\nu$ and found that $\zp$ of mass greater than $5.5\rm MeV$
is allowed from the $\Delta N_\nu$ constraint even if it is in thermal equilibrium with the neutrino bath. The grey region of Fig.\,\ref{constraint} 
depicts the constraint on $m_{\zp}$ from $\Delta N_{\nu}$. Let us note in passing that in our parameter space of interest, 
non-thermal production $\zp$ can happen for $m_{\zp} \gtrsim 40 \rm MeV$ and it does not contribute to the $\Delta N_\nu$. This is because
i) the energy density of the non-thermally produced $\zp$s are smaller compared to the thermally produced $\zp$ and they 
are non-relativistic at the time of BBN, and ii) the decay lifetime of $\zp$ in our parameter space of interest is much smaller
than $1s$. Thus the produced $\zp$ decays much earlier than SM neutrino decoupling and hence it cannot heat up the neutrino bath.   


\section{Summary and conclusions}
\label{conclusion}
In this work we have considered two sectors,  dark and visible sectors in non-standard cosmological background. 
Here, by non-standard cosmology,  we mean the presence of a new species $\phi$ which redshifts faster 
than the radiation and total energy density in the early Universe is dominated by $\phi$.
We have analyzed DM evolution in the fast expanding Universe when dark and visible sectors are not in thermal equilibrium.
To provide a detailed discussion, we have considered $U(1)_{\lmt} \otimes U(1)_X$ gauge extension of SM. 
A Dirac fermion $\x$ is the DM candidate which is charged only under $U(1)_X$ gauge symmetry. We have 
assumed tree level kinetic mixing between $U(1)_X$ and $U(1)_{\lmt}$ which generates an interaction at tree level
between $\mu$, $\tau$ flavored SM leptons with the dark gauge boson $\zp$. In standard cosmological scenario i.e. in the radiation dominated
Universe, there exists an upper limit of $\eps$ for a fixed value of $m_\x$ above which both the sectors will be in thermal equilibrium. 
In the presence of $\phi$, the upper limit of the portal coupling
is larger compared to the radiation dominated Universe. We have shown that for $n=1,\,2$, correct relic abundance of 
DM can be produced either via reannihilation or by freeze-in mechanisms
whereas for $n=3,\,4$, DM production mechanisms are freeze-in and late time annihilation. In case of late time annihilation,  excess DM is
produced from freeze-in mechanism and annihilates into $\zp$. 

Furthermore by considering the fast expanding Universe, it is possible to reconcile DM relic density and muon $(g-2)$ anomaly 
in our framework. We have also investigated the constraints from CCFR, Borexino, BABAR, WD cooling and BBN and found that $m_{\zp}-\eps$ plane is
stringently constrained from these observations. In particular, we have shown that for $n=4$, reconciliation of DM relic density for a thermally decoupled dark 
sector and muon anomalous magnetic moment is possible while satisfying the other cosmological and astrophysical constraints 
if $2\times 10^{-4} \lesssim \epsilon \lesssim 10^{-3}$, $5.5{\rm MeV} \lesssim m_{Z^\prime} \lesssim 200{\rm MeV}$, and
$1{\rm TeV} \lesssim m_\x \lesssim 10{\rm TeV}$.

\section{Acknowledgement}
SG acknowledges University Grants Commission (UGC), Government of India for providing senior research fellowship.
AT would like to acknowledge the financial support provided
by the Indian Association for the Cultivation of Science (IACS), Kolkata.  
\appendix
\section{Freeze-in production in fast expanding Universe}
\label{Appx-A}
In this appendix,  we derive approximate DM yield for pure freeze-in mechanism from 
scattering when Universe redshifts faster than radiation. To derive this,  
we need to calculate the collision term for 
$f (P_1)\bar{f} (P_2)\rightarrow \x (P_3) \bar{\x} (P_4)$ process, and it is given by,
\bea
\label{collision-term1}
\mathcal{C_E} &=& \int \prod_{i=1}^4 d\Pi_i \left(2\pi \right)^4 \delta^4\left(P_1 + P_2 - P_3 - P_4\right) 
\overline{|\mathcal{M}|^2} \, f_{1}^{\rm eq} (E_1,T)f_{2}^{\rm eq}(E_2,T) \,\,.\
\eea
Here, $P_i$, $\vec{p}_i$, and $E_i$ are four momentum, magnitude of three momentum, and energy of the $i^{th}$ species. The Lorentz
invariant phase space measure is $d \Pi_i = \dfrac{g_i d^3 \vec{p}_i}{(2\pi)^3 2 E_i}$ where $g_i$ is the internal degrees of freedom
of the $i^{th}$ species. $f_i (E_i,T)$ is the energy distribution function of the $i^{th}$ species at temperature $T$ and
$f_i (E_i,T) = \exp(-E_i/T)$. $\overline{|\mathcal{M}|^2}$ is the matrix amplitude square of $f \bar{f} \to \x \bar{\x}$ and it is averaged over
the degrees of freedom of initial and final state particles. Now, one can write $\mathcal{C_E}$ as follows
\cite{Gondolo:1990dk} 
\bea
\label{collision-term2}
\mathcal{C_E} &=& \dfrac{T}{128 \pi^4} \int_{s_{min}}^{\infty} ds\, \sqrt{s} \,\, \tilde{\sigma}\left(s\right) K_1 \left(\dfrac{\sqrt{s}}{T}\right) \,\,.\
\eea
Here, $s_{min} = {\rm Min}\left[4m_f^2, 4m_\x^2\right]$, 
$m_f$ is the mass of fermion, $K_1 (\sqrt{s}/T)$ is the modified Bessel function of second kind of order one and $\tilde{\sigma}$ is given by
$$ \tilde{\sigma} = 4 g_f^2 \left(s-4 m_f^2\right)\sigma\left(f\bar{f}\to \x\bar{\x} \right)$$

To estimate the relic density of $\x$, we calculate the annihilation cross section in the limit $s\gg m_\x^2, m_f^2$ and in this
limit $\sigma\left(\x\bar{\x} \rightarrow f \bar{f}\right) \simeq \dfrac{\alpha_X \eps^2}{16\, s}$ and one can write $\mathcal{C_E}$ as
follows
\bea
\label{collision-term}
\mathcal{C_E} &\simeq & \dfrac{\alpha_X \eps^2}{16 \pi^4}T^3m_\x K_1\left(\dfrac{2 m_\x}{T}\right)\,\,.\
\eea
To obtain final relic density of DM,  we need to solve the following Boltzmann equation, 
\bea
\label{BE-freezein}
\dfrac{dY}{dx} &=& \dfrac{1}{H(x)s(x)x} \, \mathcal{C_E}(x) \,\,.\
\eea
To get an analytical result,  we assume that the production of DM from SM bath stops at $T_{\rm fi}$ and $T_{\rm fi} \gg T_r$.
Thus, we can use the functional form of $H$ as given in Eq.\,\ref{eq:hubble_high} and the final comoving abundance of DM is given by,
\bea
\label{final-relic-freezein}
Y_\infty &\simeq & \dfrac{ 45 }{128\, \pi^7}\sqrt{\dfrac{5}{4 \pi}} 
\dfrac{\alpha_X \eps^2}{\sqrt{g_\rho \left(x_r\right)}g_{*s}} \, 
\dfrac{3\, M_{\rm Pl}}{m_\x\, x_r^{n/2}}
\,\Gamma\left(\dfrac{2+n}{4}\right)\Gamma\left(\dfrac{6+n}{4}\right) \,\,,\
\eea 
where we have neglected variation of degrees of freedom of 
entropy $g_{*s}$ with SM bath temperature $T$. Now, the relic density of DM can be calculated from $\Omega h^2 = 2.755 \times 10^8 (m_\x/1\rm GeV) Y_\infty$.

\bibliographystyle{JHEP}
\bibliography{References}
\end{document}